\documentclass[12pt]{article}

\usepackage{epsfig}

\begin{document}

\newcommand{\etal}{\em{et al.}}

\newcommand{\ra}{\rightarrow}
\newcommand{\bea}{\begin{eqnarray}}
\newcommand{\eea}{\end{eqnarray}}
\newcommand{\ubar}{\mbox{${\overline u}$}}
\newcommand{\dbar}{\mbox{${\overline d}$}}
\newcommand{\sbar}{\mbox{${\overline s}$}}
\newcommand{\cbar}{\mbox{${\overline c}$}}
\newcommand{\bbar}{\mbox{${\overline b}$}}
\newcommand{\bbbar}{\mbox{$b\bbar$}}
\newcommand{\ccbar}{\mbox{$c\cbar$}}
\def\ccbar{\mbox{${c\cbar}$}}
\newcommand{\qbar}{\mbox{${\overline q}$}}
\newcommand{\qqbar}{\mbox{$q\qbar$}}
\newcommand{\aip}{\mbox{$\langle {\rm IP}\rangle$}}
\newcommand{\sip}{\mbox{$\sigma_{_{IP}}$}}
\newcommand{\zmumu}{\mbox{${Z^0\ra\mu^+\mu^-}$}}
\newcommand{\zbb}{\mbox{${Z^0\ra b\bbar}$}}
\newcommand{\zcbar}{\mbox{${Z^0\ra c\cbar}$}}
\newcommand{\zuds}{\mbox{${Z^0\ra u\ubar+c\cbar+s\sbar$}}}
\newcommand{\zudsc}{\mbox{${Z^0\ra udsc}$}}
\newcommand{\zqq}{\mbox{${Z^0\ra hadrons}$}}
\newcommand{\gzcc}{\mbox{${\Gamma(\zcbar)}$}}
\newcommand{\zee}{\mbox{${Z^0\ra e^{+}e^-}$}}
\newcommand{\ee}{\mbox{${e^{+}e^-}$}}
\newcommand{\ep}{\mbox{${e^{+}e^-}$}}
\newcommand{\effb}{\mbox{${\epsilon_b}$}}
\newcommand{\efbb}{\mbox{${\epsilon_b^2}$}}
\newcommand{\effc}{\mbox{${\epsilon_c}$}}
\newcommand{\efcc}{\mbox{${\epsilon_c^2}$}}
\newcommand{\effu}{\mbox{${\epsilon_{uds}}$}}
\newcommand{\efuu}{\mbox{${\epsilon_{uds}^2}$}}
\newcommand{\lamb}{\mbox{${\lambda_b}$}}
\newcommand{\lamc}{\mbox{${\lambda_c}$}}
\newcommand{\purb}{\mbox{${\Pi_b}$}}
\newcommand{\Mtag}{\mbox{${\cal M}$}}
\newcommand{\glubb}{\mbox{${g\ra{b\overline{b}}}$}}
\newcommand{\glucc}{\mbox{${g\ra{c\overline{c}}}$}}
\newcommand{\gluQQ}{\mbox{${g\ra{Q\overline{Q}}}$}}
\newcommand{\ycut}{\mbox{$y_{\mbox{\tiny cut}}$}}

\newcommand{\gev}{\ensuremath{\mathrm{\,Ge\kern -0.1em V}}}
\newcommand{\mev}{\ensuremath{\mathrm{\,Me\kern -0.1em V}}}
\newcommand{\gevc}{\ensuremath{{\mathrm{\,Ge\kern -0.1em V\!/}c}}}
\newcommand{\mevc}{\ensuremath{{\mathrm{\,Me\kern -0.1em V\!/}c}}}
\newcommand{\gevcc}{\ensuremath{{\mathrm{\,Ge\kern -0.1em V\!/}c^2}}}
\newcommand{\mevcc}{\ensuremath{{\mathrm{\,Me\kern -0.1em V\!/}c^2}}}
\def\cm   {\ensuremath{\rm \,cm}}
\def\mm   {\ensuremath{\rm \,mm}}
\def\mum  {\ensuremath{\,\mu\rm m}}

\newcommand{\Z}{\mbox{$Z^0$}}

\newcommand{\plb}{Phys. Lett.}
\newcommand{\npb}{Nucl. Phys.}
\newcommand{\rmp}{Rev. Mod. Phys.}
\newcommand{\prl}{Phys. Rev. Lett.}
\newcommand{\prd}{Phys. Rev.}
\newcommand{\zpc}{Z. Phys.}


\newcommand{\RBF}{0.21604}
\newcommand{\DRBSTAT}{0.00098}
\newcommand{\DRBSYS}{0.00073}
\newcommand{\DRBRC}{0.00012}
\newcommand{\PRB}{97.9\%}
\newcommand{\EFRB}{62.0\%}
\newcommand{\DEFRB}{0.24\%}
\newcommand{\MCLAMB}{-0.02\%}

\newcommand{\RCF}{0.1744}
\newcommand{\DRCSTAT}{0.0031}
\newcommand{\DRCSYS}{0.0020}
\newcommand{\DRCRB}{0.0006}
\newcommand{\PRC}{84.5\%}
\newcommand{\EFRC}{17.9\%}
\newcommand{\DEFRC}{0.6\%}

\thispagestyle{empty}
\begin{flushright}
{\renewcommand{\baselinestretch}{.75}
  SLAC--PUB--9941\\
February, 2005\\
}
\end{flushright}

\vskip 0.2truecm
\begin{center}
{\large\bf
 Measurement of the branching ratios of the $Z^0$ into heavy quarks$^*$
}
\end{center}
%
%
%
\begin{center}
\def\iAOMORI{$^{(1)}$}
\def\iBRUN{$^{(2)}$}
\def\iBU{$^{(3)}$}
\def\iCOLO{$^{(4)}$}
\def\iCSU{$^{(5)}$}
\def\iFERR{$^{(6)}$}
\def\iFRAS{$^{(7)}$}
\def\iJHU{$^{(8)}$}
\def\iLBL{$^{(9)}$}
\def\iMASS{$^{(10)}$}
\def\iMISSI{$^{(11)}$}
\def\iMIT{$^{(12)}$}
\def\iMOSCOW{$^{(13)}$}
\def\iNAGO{$^{(14)}$}
\def\iOREG{$^{(15)}$}
\def\iOXF{$^{(16)}$}
\def\iPERU{$^{(17)}$}
\def\iRAL{$^{(18)}$}
\def\iRUTG{$^{(19)}$}
\def\iSLAC{$^{(20)}$}
\def\iSOONG{$^{(21)}$}
\def\iTENN{$^{(22)}$}
\def\iTOHO{$^{(23)}$}
\def\iUCSB{$^{(24)}$}
\def\iUCSC{$^{(25)}$}
\def\iVAND{$^{(26)}$}
\def\iWASH{$^{(27)}$}
\def\iWISC{$^{(28)}$}
\def\iYALE{$^{(29)}$}

\baselineskip=.75\baselineskip
\mbox{Koya Abe\unskip,\iTOHO}
\mbox{Kenji Abe\unskip,\iNAGO}
\mbox{T. Abe\unskip,\iSLAC}
\mbox{I. Adam\unskip,\iSLAC}
\mbox{H. Akimoto\unskip,\iSLAC}
\mbox{D. Aston\unskip,\iSLAC}
\mbox{K.G. Baird\unskip,\iMASS}
\mbox{C. Baltay\unskip,\iYALE}
\mbox{H.R. Band\unskip,\iWISC}
\mbox{T.L. Barklow\unskip,\iSLAC}
\mbox{J.M. Bauer\unskip,\iMISSI}
\mbox{G. Bellodi\unskip,\iOXF}
\mbox{R. Berger\unskip,\iSLAC}
\mbox{G. Blaylock\unskip,\iMASS}
\mbox{J.R. Bogart\unskip,\iSLAC}
\mbox{G.R. Bower\unskip,\iSLAC}
\mbox{J.E. Brau\unskip,\iOREG}
\mbox{M. Breidenbach\unskip,\iSLAC}
\mbox{W.M. Bugg\unskip,\iTENN}
\mbox{D. Burke\unskip,\iSLAC}
\mbox{T.H. Burnett\unskip,\iWASH}
\mbox{P.N. Burrows\unskip,\iOXF}
\mbox{A. Calcaterra\unskip,\iFRAS}
\mbox{R. Cassell\unskip,\iSLAC}
\mbox{A. Chou\unskip,\iSLAC}
\mbox{H.O. Cohn\unskip,\iTENN}
\mbox{J.A. Coller\unskip,\iBU}
\mbox{M.R. Convery\unskip,\iSLAC}
\mbox{V. Cook\unskip,\iWASH}
\mbox{R.F. Cowan\unskip,\iMIT}
\mbox{G. Crawford\unskip,\iSLAC}
\mbox{C.J.S. Damerell\unskip,\iRAL}
\mbox{M. Daoudi\unskip,\iSLAC}
\mbox{N. de Groot\unskip,\iRAL}
\mbox{R. de Sangro\unskip,\iFRAS}
\mbox{D.N. Dong\unskip,\iMIT}
\mbox{M. Doser\unskip,\iSLAC}
\mbox{R. Dubois\unskip,}
\mbox{I. Erofeeva\unskip,\iMOSCOW}
\mbox{V. Eschenburg\unskip,\iMISSI}
\mbox{E. Etzion\unskip,\iWISC}
\mbox{S. Fahey\unskip,\iCOLO}
\mbox{D. Falciai\unskip,\iFRAS}
\mbox{J.P. Fernandez\unskip,\iUCSC}
\mbox{K. Flood\unskip,\iMASS}
\mbox{R. Frey\unskip,\iOREG}
\mbox{E.L. Hart\unskip,\iTENN}
\mbox{K. Hasuko\unskip,\iTOHO}
\mbox{S.S. Hertzbach\unskip,\iMASS}
\mbox{M.E. Huffer\unskip,\iSLAC}
\mbox{X. Huynh\unskip,\iSLAC}
\mbox{M. Iwasaki\unskip,\iOREG}
\mbox{D.J. Jackson\unskip,\iRAL}
\mbox{P. Jacques\unskip,\iRUTG}
\mbox{J.A. Jaros\unskip,\iSLAC}
\mbox{Z.Y. Jiang\unskip,\iSLAC}
\mbox{A.S. Johnson\unskip,\iSLAC}
\mbox{J.R. Johnson\unskip,\iWISC}
\mbox{R. Kajikawa\unskip,\iNAGO}
\mbox{M. Kalelkar\unskip,\iRUTG}
\mbox{H.J. Kang\unskip,\iRUTG}
\mbox{R.R. Kofler\unskip,\iMASS}
\mbox{R.S. Kroeger\unskip,\iMISSI}
\mbox{M. Langston\unskip,\iOREG}
\mbox{D.W.G. Leith\unskip,\iSLAC}
\mbox{V. Lia\unskip,\iMIT}
\mbox{C. Lin\unskip,\iMASS}
\mbox{G. Mancinelli\unskip,\iRUTG}
\mbox{S. Manly\unskip,\iYALE}
\mbox{G. Mantovani\unskip,\iPERU}
\mbox{T.W. Markiewicz\unskip,\iSLAC}
\mbox{T. Maruyama\unskip,\iSLAC}
\mbox{A.K. McKemey\unskip,\iBRUN}
\mbox{R. Messner\unskip,\iSLAC}
\mbox{K.C. Moffeit\unskip,\iSLAC}
\mbox{T.B. Moore\unskip,\iYALE}
\mbox{M. Morii\unskip,\iSLAC}
\mbox{D. Muller\unskip,\iSLAC}
\mbox{V. Murzin\unskip,\iMOSCOW}
\mbox{S. Narita\unskip,\iTOHO}
\mbox{U. Nauenberg\unskip,\iCOLO}
\mbox{H. Neal\unskip,\iYALE}
\mbox{G. Nesom\unskip,\iOXF}
\mbox{N. Oishi\unskip,\iNAGO}
\mbox{D. Onoprienko\unskip,\iTENN}
\mbox{L.S. Osborne\unskip,\iMIT}
\mbox{R.S. Panvini\unskip,\iVAND}
\mbox{C.H. Park\unskip,\iSOONG}
\mbox{I. Peruzzi\unskip,\iFRAS}
\mbox{M. Piccolo\unskip,\iFRAS}
\mbox{L. Piemontese\unskip,\iFERR}
\mbox{R.J. Plano\unskip,\iRUTG}
\mbox{R. Prepost\unskip,\iWISC}
\mbox{C.Y. Prescott\unskip,\iSLAC}
\mbox{B.N. Ratcliff\unskip,\iSLAC}
\mbox{J. Reidy\unskip,\iMISSI}
\mbox{P.L. Reinertsen\unskip,\iUCSC}
\mbox{L.S. Rochester\unskip,\iSLAC}
\mbox{P.C. Rowson\unskip,\iSLAC}
\mbox{J.J. Russell\unskip,\iSLAC}
\mbox{O.H. Saxton\unskip,\iSLAC}
\mbox{T. Schalk\unskip,\iUCSC}
\mbox{B.A. Schumm\unskip,\iUCSC}
\mbox{J. Schwiening\unskip,\iSLAC}
\mbox{V.V. Serbo\unskip,\iSLAC}
\mbox{G. Shapiro\unskip,\iLBL}
\mbox{N.B. Sinev\unskip,\iOREG}
\mbox{J.A. Snyder\unskip,\iYALE}
\mbox{H. Staengle\unskip,\iCSU}
\mbox{A. Stahl\unskip,\iSLAC}
\mbox{P. Stamer\unskip,\iRUTG}
\mbox{H. Steiner\unskip,\iLBL}
\mbox{D. Su\unskip,\iSLAC}
\mbox{F. Suekane\unskip,\iTOHO}
\mbox{A. Sugiyama\unskip,\iNAGO}
\mbox{A. Suzuki\unskip,\iNAGO}
\mbox{M. Swartz\unskip,\iJHU}
\mbox{F.E. Taylor\unskip,\iMIT}
\mbox{J. Thom\unskip,\iSLAC}
\mbox{E. Torrence\unskip,\iMIT}
\mbox{T. Usher\unskip,\iSLAC}
\mbox{J. Va'vra\unskip,\iSLAC}
\mbox{R. Verdier\unskip,\iMIT}
\mbox{D.L. Wagner\unskip,\iCOLO}
\mbox{A.P. Waite\unskip,\iSLAC}
\mbox{S. Walston\unskip,\iOREG}
\mbox{A.W. Weidemann\unskip,\iTENN}
\mbox{E.R. Weiss\unskip,\iWASH}
\mbox{J.S. Whitaker\unskip,\iBU}
\mbox{S.H. Williams\unskip,\iSLAC}
\mbox{S. Willocq\unskip,\iMASS}
\mbox{R.J. Wilson\unskip,\iCSU}
\mbox{W.J. Wisniewski\unskip,\iSLAC}
\mbox{J.L. Wittlin\unskip,\iMASS}
\mbox{M. Woods\unskip,\iSLAC}
\mbox{T.R. Wright\unskip,\iWISC}
\mbox{R.K. Yamamoto\unskip,\iMIT}
\mbox{J. Yashima\unskip,\iTOHO}
\mbox{S.J. Yellin\unskip,\iUCSB}
\mbox{C.C. Young\unskip,\iSLAC}
\mbox{H. Yuta\unskip.\iAOMORI}
\it
  \vskip \baselineskip                   
  \centerline{(The SLD Collaboration)}   
  \vskip \baselineskip
  \baselineskip=.75\baselineskip   
\iAOMORI
  Aomori University, Aomori, 030 Japan, \break
\iBRUN
  Brunel University, Uxbridge, Middlesex, UB8 3PH United Kingdom, \break
\iBU
  Boston University, Boston, Massachusetts 02215, \break
\iCOLO
  University of Colorado, Boulder, Colorado 80309, \break
\iCSU
  Colorado State University, Ft. Collins, Colorado 80523, \break
\iFERR
  INFN Sezione di Ferrara and Universita di Ferrara, I-44100 Ferrara, Italy,
\break
\iFRAS
  INFN Laboratori Nazionali di Frascati, I-00044 Frascati, Italy, \break
\iJHU
  Johns Hopkins University,  Baltimore, Maryland 21218-2686, \break
\iLBL
  Lawrence Berkeley Laboratory, University of California, Berkeley, California
94720, \break
\iMASS
  University of Massachusetts, Amherst, Massachusetts 01003, \break
\iMISSI
  University of Mississippi, University, Mississippi 38677, \break
\iMIT
  Massachusetts Institute of Technology, Cambridge, Massachusetts 02139, \break
\iMOSCOW
  Institute of Nuclear Physics, Moscow State University, 119899 Moscow, Russia,
\break
\iNAGO
  Nagoya University, Chikusa-ku, Nagoya, 464 Japan, \break
\iOREG
  University of Oregon, Eugene, Oregon 97403, \break
\iOXF
  Oxford University, Oxford, OX1 3RH, United Kingdom, \break
\iPERU
  INFN Sezione di Perugia and Universita di Perugia, I-06100 Perugia, Italy,
\break
\iRAL
  Rutherford Appleton Laboratory, Chilton, Didcot, Oxon OX11 0QX United Kingdom,
\break
\iRUTG
  Rutgers University, Piscataway, New Jersey 08855, \break
\iSLAC
  Stanford Linear Accelerator Center, Stanford University, Stanford, California
94309, \break
\iSOONG
  Soongsil University, Seoul, Korea 156-743, \break
\iTENN
  University of Tennessee, Knoxville, Tennessee 37996, \break
\iTOHO
  Tohoku University, Sendai, 980 Japan, \break
\iUCSB
  University of California at Santa Barbara, Santa Barbara, California 93106,
\break
\iUCSC
  University of California at Santa Cruz, Santa Cruz, California 95064, \break
\iVAND
  Vanderbilt University, Nashville,Tennessee 37235, \break
\iWASH
  University of Washington, Seattle, Washington 98105, \break
\iWISC
  University of Wisconsin, Madison,Wisconsin 53706, \break
\iYALE
  Yale University, New Haven, Connecticut 06511. \break
\rm
%

\end{center}

\vskip 1truecm

\eject
\begin{center}
{\bf ABSTRACT }
\end{center}

\noindent
We measure the hadronic branching ratios
of the \Z\ boson into heavy quarks: 
$R_b=\Gamma_{Z^0\rightarrow b\overline{b}}/\Gamma_{Z^0\rightarrow hadrons}$
and
$R_c=\Gamma_{Z^0\rightarrow c\overline{c}}/\Gamma_{Z^0\rightarrow hadrons}$
using a multi-tag technique. The measurement was performed
using about 400,000 hadronic \Z\ events recorded in the SLD experiment 
at SLAC between 1996 and 1998.  The small and stable SLC beam spot 
and the CCD-based vertex detector were used to reconstruct 
bottom and charm hadron decay vertices with high efficiency and purity, 
which enables us to measure most efficiencies from data. We obtain,
$$R_b=\RBF\pm\DRBSTAT(stat.)\pm\DRBSYS(syst.)\mp\DRBRC(R_c)$$
and, 
$$R_c=\RCF\pm\DRCSTAT(stat.)\pm\DRCSYS(syst.)\mp\DRCRB(R_b)$$

\vfill
{\footnotesize
$^*$ Work supported by Department of Energy contract DE-AC03-76SF00515 (SLAC).}

\eject

\section{Introduction}
 
\noindent
The dominant production of $Z^0$ decays, with large statistics, 
at \ee\ experiments operating on the \Z\ peak, provides a 
unique opportunity for probing electroweak interactions at high 
precision. The democratic production of all fermion flavors 
in \Z\ decays with a clean initial state allows particularly
sensitive tests of the Standard Model (SM) through measurements
of
$R_b=\Gamma_{Z^0\rightarrow b\overline{b}}/\Gamma_{Z^0\rightarrow hadrons}$
and
$R_c=\Gamma_{Z^0\rightarrow c\overline{c}}/\Gamma_{Z^0\rightarrow hadrons}$,
the heavy quark production fractions in hadronic \Z\ decays.
The $b$ and $c$ quarks are the heaviest charge 1/3 and charge 2/3 
quarks, respectively, that are accessible at the \Z\ energy.
The $R_b$ measurement has been traditionally a focus of attention  
as it is not only sensitive to the heavy top quark mass, but also 
widely regarded as a promising window for detecting new physics 
through radiative corrections to the \zbb\ coupling.   
More generally, any unexpected difference in quark coupling of 
one flavor compared with other flavors could be a vital clue 
toward a solution to the puzzle of fermion family degeneracy.    

To take full advantage of the large \Z\ decay samples as 
an effective flavor physics arena, the advance in vertex detector 
technologies has played a key role for flavor identification.  
Unprecedented performance in $b$ quark tagging has made it 
possible to achieve $R_b$ measurements~\cite{leprb}\cite{sldrb}
at below 1\% precision. Our previous $R_b$ measurement 
\cite{sldrb} has demonstrated the effectiveness of the 
combination of high resolution vertexing and the small and stable 
interaction point at the SLC, which pointed the way for very high 
purity $b$-tags and led to the reduction of systematic 
uncertainties for $R_b$ measurements in general. In this 
paper we present an updated $R_b$ measurement from SLD using 
$\sim$2.5 times more statistics than our previous publication
and with an upgraded CCD pixel vertex detector, which improves the
$b$-tag efficiency to approximately a factor of two higher than
$b$-tags used in existing measurements at similar $b$-purity.       
  
The existing measurements of the charm branching ratio 
$R_c$~\cite{leprc} are considerably less precise than the $R_b$ 
measurements. These determinations of $R_c$
rely on a number of methods to identify charm jets that all have
certain disadvantages. Exclusive reconstruction of charmed mesons
suffers from a small branching fraction and introduces a dependence
on the actual value of this branching fraction and on the production
fractions of the charm hadrons. Inclusive reconstruction
using leptons or slow pions suffer from low purities and again a dependence
on branching fractions. Due to the superb performance of the SLD vertex
detector we are able to introduce an inclusive charm tag,
based on lifetime information in a similar way as the $b$ tag, which
combines high efficiency with good purity. Moreover, the tagging 
efficiencies are measured from data and result in a minimal reliance 
on Monte Carlo simulation and a small systematic uncertainty. 
This leads to a new determination of $R_c$ with much improved overall 
precision than the existing measurements.

\section{Apparatus and Hadronic Event Selection}
\label{sec:det-hadsel}

\noindent
This analysis is based on approximately 400,000 hadronic events produced in 
\ep\ annihilations at a mean center-of-mass energy of $\sqrt{s}=91.28$ GeV
at the SLAC Linear Collider (SLC), and recorded in the SLC Large Detector
(SLD) between 1996 and 1998. 
A general description of the SLD can be found elsewhere~\cite{sld}.
The trigger and initial selection criteria for hadronic $Z^0$ decays are 
described in Ref.~\cite{vxd2RbPrd}.
The Central Drift Chamber (CDC)~\cite{cdc} and the upgraded Vertex 
Detector (VXD3)~\cite{vxd}, inside a uniform axial magnetic field of 0.6T,
provide the momentum measurements of charged tracks and precision vertex 
information near the interaction point (IP), which are central to this 
analysis.  
The energies of clusters measured in the Liquid Argon Calorimeter~\cite{lac}
are used for event selection and calculation of the event thrust axis.  

SLD uses a coordinate system with the $z$-axis parallel to the beam direction
and $x$ and $y$ respectively are the horizontal and vertical coordinates 
perpendicular to the $z$-axis. The polar angle $\theta$ is measured with 
respect to the $z$-axis and the azimuthal angle $\phi$ is the angle with the
$x$-axis in the $xy$ plane.

The CDC and VXD3 give a combined momentum resolution of
$\sigma_{p_{\perp}}/p_{\perp}$ = $0.01 \oplus 0.0026p_{\perp}$,
where $p_{\perp}$ is the track momentum transverse to the beam axis in
\gevc. The VXD3 consists of 3 barrels of Charged Coupled Devices (CCD)
at radii of 2.7, 3.8 and 4.8 centimeters from the beam line, with 
3-hit solid angle coverage up to $|\cos\theta|$=0.85. The CCD pixels 
are cubic active volumes of 20\mum\ on each side, which provide 
precise 3D spatial hits. This high granularity in 3D space
enable our track finding algorithm to limit the 
hit mis-assignment rate to only 0.2\%.   
The spatial resolution on the hit cluster centroid achieved after a 
track based CCD alignment \cite{VXD3alignment} is $\sim$4\mum\ in both 
azimuthal and $z$ directions. The resultant tracking resolution for high 
momentum tracks, as measured from the miss distances of two tracks near 
the IP in \zmumu\ events, is 7.7\mum\ in $r\phi$ and 9.6\mum\ in $rz$. 
The multiple scattering contribution to the track impact parameter 
resolution can be approximately expressed as 
$\frac{33}{p \sin^{3/2}\theta}$~\mum. These numbers
are roughly a factor of two better than a typical vertex detector 
at LEP. This resolution advantage combined with the small and stable 
SLC IP information is crucial in establishing the feasibility of
the measurement techniques, in particular for the case of the inclusive charm 
tagging for the $R_c$ measurement.     

For the purpose of estimating the efficiencies and purities of the 
event selection and flavor tagging procedures, we use a detailed Monte Carlo
(MC) simulation of the detector.
The JETSET 7.4~\cite{jetset} event generator is used, with parameter
values tuned to hadronic \ep\ annihilation data~\cite{tune},
combined with a simulation of the SLD based on GEANT 3.21~\cite{geant}.
The simulations of heavy hadron production and charm decays
are described in \cite{vxd2RbPrd}. The $B$ decay simulation is adapted
from the CLEO QQ MC with additional tuning by SLD (see appendix B of 
\cite{chou-thesis}) to match a wider range of $\Upsilon(4S)$ $B$ decay data.   

For the hadronic event selection, we use a set of well-measured tracks, 
consistent with originating from within the beam pipe radius of 2.5 cm.
The selected events have at least 7 tracks with $p_\perp >$0.2\gevc\ 
and within 5\cm\ from the interaction 
point along the beam axis. There are at least 3 tracks with two or 
more VXD hits. The event visible energy $E_{vis}$, calculated from 
charged tracks with the charged pion mass assigned, must be at least 
18 GeV. To ensure the events are well contained within the detector 
fiducial volume, we require the thrust-axis polar angle w.r.t. the beamline, 
$\theta_T$, calculated using calorimeter clusters, to
be within $|\cos\theta_T|<$0.7. We only include events with $\leq$3 jets
for the analysis, where the jets are reconstructed from tracks
using the JADE algorithm \cite{JADE} 
with $y_{cut}=0.02$. This last requirement
is to reject higher jet multiplicity events where the division of
events into two hemispheres is no longer reliable for partitioning
the events with one heavy hadron in each hemisphere. 

The selected event sample contains 191,770 events from the 1997-98 runs 
and 29,996 events from the 1996 run. We analyze the two data samples 
separately because the small 1996 sample, recorded during the early VXD3 
commissioning period, has a lower overall VXD hit efficiency due 
to electronics problems and some radiation damage. This effect
was not present in the 97-98 period when the VXD3 ran at $\sim$10 
degrees colder and the operation of the electronics was much more stable. 

The estimated background 
in the hadronic event sample is negligible. The Monte Carlo 
\zqq\ event statistics used in the analysis have MC:data event ratios 
of 4:1, 15:1, and 22:1 for light flavor quarks ($uds$), 
\ccbar, and \bbbar\ events, respectively.

\section{Heavy Flavor Tagging}

Decays of $Z^0$ bosons into charm and bottom quarks can be distinguished 
from those into light flavors ($u$,$d$ and $s$) by searching for 
the heavy hadron decay vertices displaced from the event interaction point (IP).
Because they are produced with high energy and have long lifetimes, 
heavy hadrons generally travel distances of millimeters before decaying. 
In a jet from a light quark the tracks will appear to come 
from one point in
space, the event IP. In a charm jet some of the
tracks may not point back to the IP, and if the charmed hadron decays into
more than one charged particle there will be a secondary vertex (SV) in
addition to the IP.  Bottom jets will also exhibit secondary vertices, 
and if there are sufficient particles produced at the $b$ and $c$ 
decay points it is possible to find more than one displaced vertex. 
The reconstructed secondary vertices and their associated kinematic 
properties serve as the primary basis for flavor tagging for both 
\bbbar\ and \ccbar\ events in this analysis.    

\subsection{IP Reconstruction}

To search for displaced vertices the position of the IP must be 
precisely known.  In each event the position of the IP in the plane transverse
to the beam axis is determined by fitting all tracks
that are compatible with coming from the IP to
a common vertex. Because the SLC luminous region is small and stable in
the $xy$ plane, sets of 30 time sequential hadronic events are averaged 
to
obtain a more precise determination of the $xy$ IP position
(details in appendix A of \cite{chou-thesis}). 
The $xy$ IP resolution is measured from the impact parameter 
distributions of \zmumu\ events, deconvolved from the track impact 
parameter resolution measured from the two track miss distance, to 
obtain an IP resolution of 3.2\mum.        

Because the SLC luminous region is larger in $z$ (700 $\mu$m), the 
$z$ position of the IP must be found event-by-event. Tracks with VXD 
hits are extrapolated to their point of closest approach (POCA) in $xy$ 
to the precisely determined transverse IP position. Tracks with impact
parameters of more than 500\mum\ or 3 times the error
($\sigma$) from the IP are excluded, and the
$z$ position of the IP is taken from the median $z$ at POCA of the
remaining tracks.  The resolution of this method is found from the
Monte Carlo simulation to be 10/11/17 \mum\ for $uds$/\ccbar/\bbbar\ 
events.

\subsection{Secondary Vertex Reconstruction}

Secondary vertices are found using a topological algorithm \cite{zvnim}.  
This method searches for space points of large track density in 3 
dimensions.  
Each track is parameterized by a Gaussian probability density tube 
$f(\vec{r})$ with a width equal to the uncertainty in the track position 
at its POCA to the IP, $\vec{r}_0$:

\begin{equation}
f(\vec{r})=\exp\left\{-\frac{1}{2}\left[\left(\frac{x-(x_0+y^2\kappa)}
{\sigma_1}\right)^2 +\left(\frac{z-(z_0+y\tan\lambda)}
{\sigma_2}\right)^2\right]\right\}.
\end{equation}

The first term is a parabolic approximation to the track's circular
trajectory in the $xy$ plane, where $\kappa$ is a function of the
track's charge and transverse momentum and of the SLD magnetic field.
The second term represents the linear trajectory of the track in the $rz$
plane, where $\lambda$ is the track dip angle from the vertical. The
$\sigma$ parameters are the uncertainties in the track positions after
extrapolation to $\vec{r}_0$ for the two projections. The function
$f_i(\vec{r})$ is formed for each track $i$ under consideration and used to
construct the vertex probability function $V(\vec{r})$.  Also included is
$f_0(\vec{r})$, a $7\times 7\times 20$ $\mu$m $(x\times y\times
z)$ Gaussian ellipsoid centered at the IP
position.

 \begin{equation}
V(\vec{r})=\sum_i f_i(\vec{r})-\frac{\sum_i f_i^2(\vec{r})}
{\sum_i f_i(\vec{r})}
\end{equation}

Secondary vertices are found by searching for local maxima in
$V(\vec{r})$ that are well-separated from the peak at the IP position.  
The tracks whose density functions contribute to such a local maximum 
are then identified as originating from a secondary vertex (SV).

\begin{figure}
\caption{Distributions of seed vertex selection variables: 
(a) distance from IP $D$, 
(b) normalized distance from IP $D/\sigma_{D}$, 
(c) angle between flight direction and vertex momentum $\phi_{PD}$, 
(d) neural network output $y_{vtx}$.  
A good vertex contains only heavy hadron decay tracks.  
The arrow indicates the accepted region.}
\label{f:vtxnn}
\vspace{.2in}
\begin{center}
\epsfig{figure=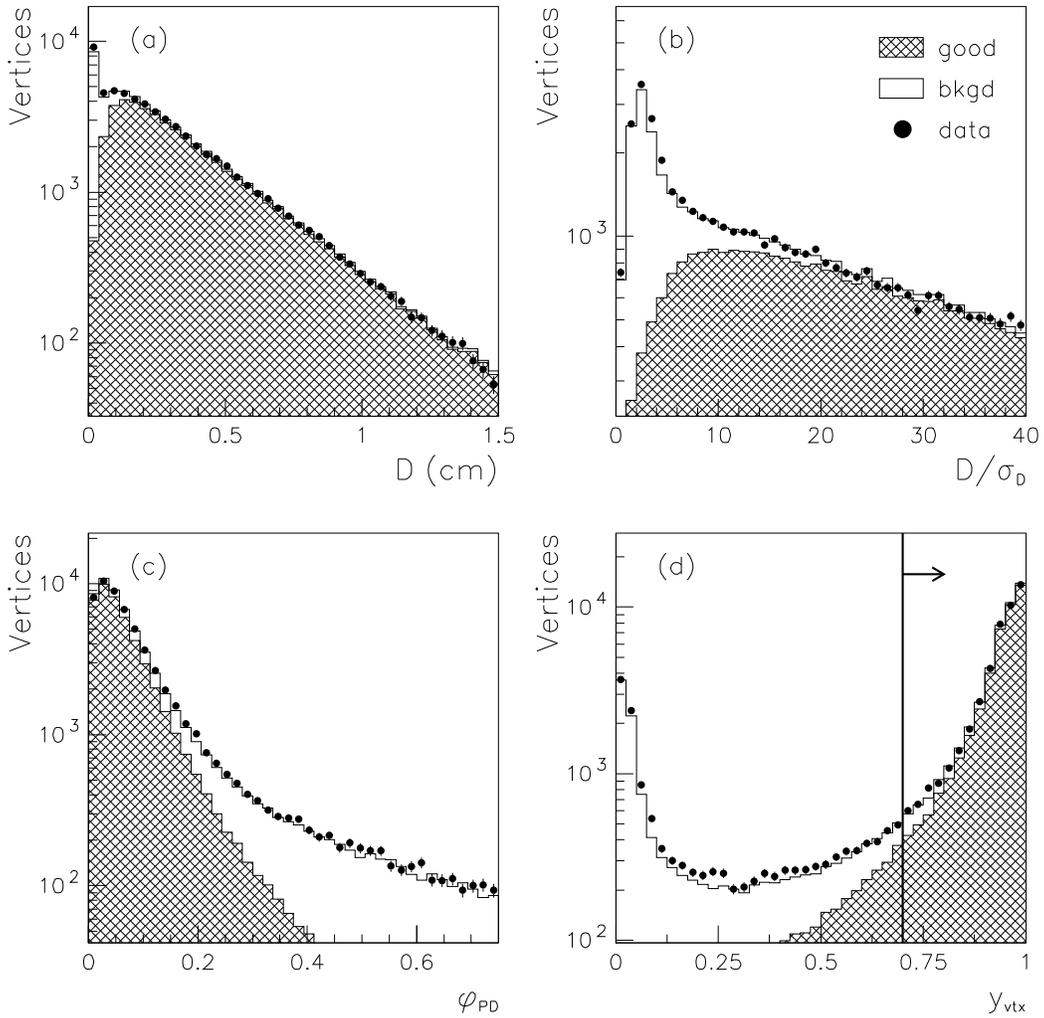,width=\textwidth}
\end{center}
\end{figure}

A loose set of cuts is applied to tracks used for secondary vertex
reconstruction.  Tracks are required to have $\ge 2$ VXD hits and
$p_{\perp}>250$\mevc.  Tracks with 3D impact parameter $>3$ mm or
consistent with originating from a $K_s^0$ or $\Lambda$ decay, or
$\gamma$ conversion 
are also removed. Each event is divided into two hemispheres using the
thrust axis, and the vertexing procedure is performed in each using only
the tracks in that hemisphere. The identified vertices are required to be
within 2.3 cm of the center of the beam pipe to remove false vertices from
interactions with the detector material.  A cut on the secondary vertex
invariant mass $M$ of \mbox{$|M-M_{K^0_S}|<0.015$ GeV/$c^2$} removes any $K^0_S$
decays that survived the track cuts. The remaining vertices are then
passed through a neural network \cite{wright-thesis} 
to improve the background rejection further. 
The input variables are the flight distance from the IP to
the vertex ($D$), that distance normalized by its error ($D/\sigma_D$), and
the angle $\phi_{PD}$ between the flight direction $\vec{D}$ and the 
total momentum vector of the vertex $\vec{P}$. These quantities are shown
in Figure \ref{f:vtxnn}, along with the output value of the neural network
($y_{vtx}$).  A good vertex in simulation is defined as one which contains only tracks
from heavy hadron decays, with no tracks originating from the IP, strange
particle decays, or other sources.  Vertices with $y_{vtx}>0.7$ are
retained.  At least one secondary vertex passing this cut is found in
72.7\% of bottom, 28.2\% of charm, and 0.41\% of light quark event
hemispheres in the Monte Carlo.  Around 16\% of the hemispheres in $b$
events have more than one selected secondary vertex.

\subsection{Track Attachment}

\begin{figure}
\caption{Schematic illustration of the quantities used in the 
track-attachment procedure described in the text (not to scale).}
\label{f:loverd}
\vspace{.2in}
\begin{center}
\epsfig{figure=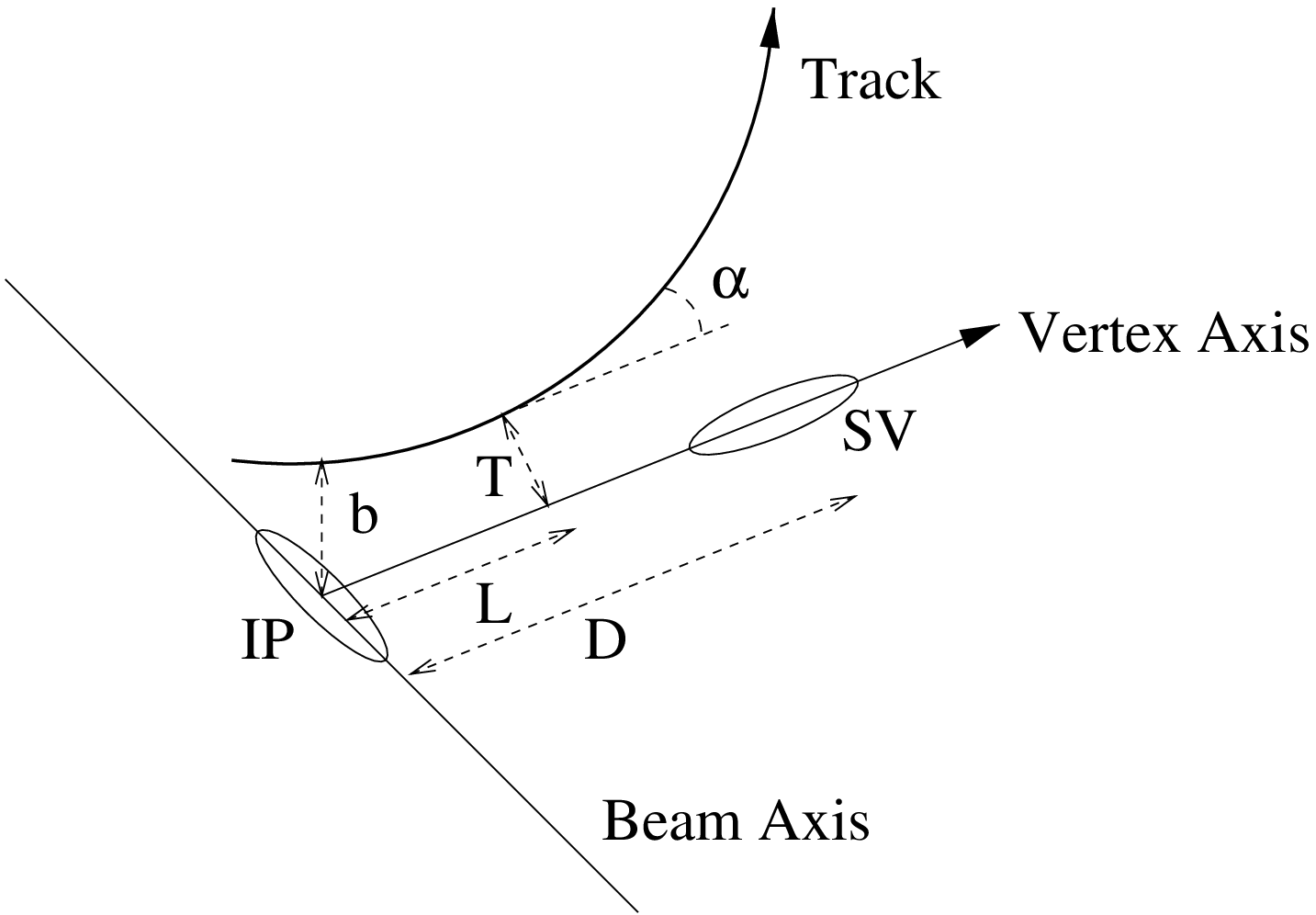,width=0.6\textwidth}
\end{center}
\end{figure}

\begin{figure}
\caption{Distributions of the cascade track selection variables
described in the text: 
(a) $T$, (b) $L$, (c) $L/D$, (d) $\alpha$, (e) $b/\sigma_b$, 
(f) neural network output $y_{trk}$.  
A good track is one which originates from a heavy hadron decay. 
The arrow indicates the accepted region.}
\label{f:trknn}
\vspace{.2in}
\begin{center}
\epsfig{figure=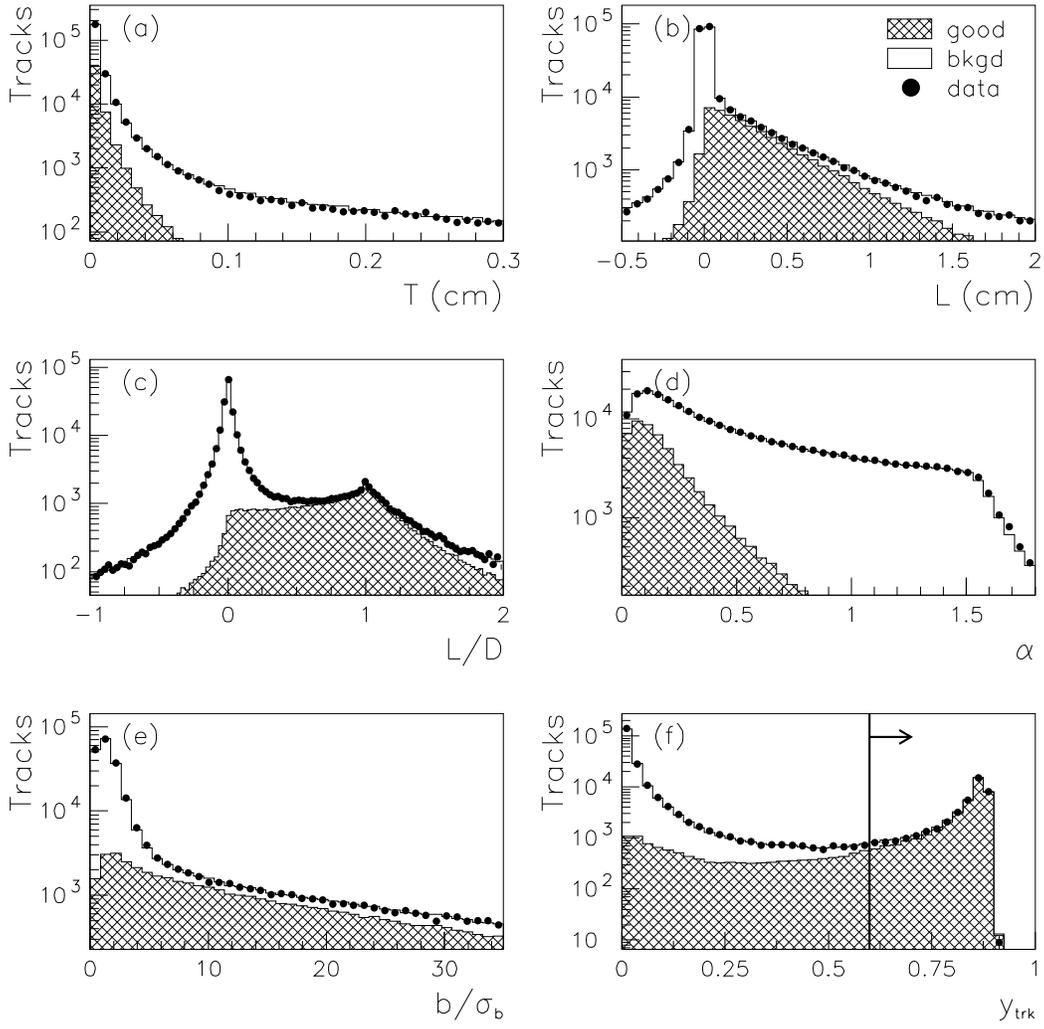,width=\textwidth}
\end{center}
\end{figure}

Due to the cascade nature of $b$ hadron decays, tracks from
the heavy hadron may not all originate from the same space point.  
Therefore, a process of attaching tracks to the secondary vertex (SV) 
has been developed to recover this information using a second neural network.  
The first four inputs are defined at the point of closest approach of the
track to the axis joining the secondary vertex to the IP. They are
the transverse distance from the track to that axis ($T$), the
distance from the IP along that axis to the POCA ($L$), that distance
divided by the flight distance of the SV from the IP ($L/D$), and the
angle of the track to the IP-SV axis ($\alpha$). The last input is the 3D
impact parameter of the track to the IP normalized by its error
($b/\sigma_b$). These quantities are shown schematically in
Figure~\ref{f:loverd}. The distributions are shown in
Figure~\ref{f:trknn}, along with the neural network output value
($y_{trk}$).  The network is trained to accept only tracks which come from
a $b$ or $c$ hadron decay, and to reject tracks from the IP or from
strange particle decays or detector interaction products. If more than one
secondary vertex was found in the hemisphere the attachment procedure is
tried for each track-SV combination.  Tracks with $y_{trk}>0.6$ are added
to the list of secondary vertex tracks. This value is chosen to minimize
the number of fake tracks being attached to charm vertices, which than
could mimic a $b$ decay.

\subsection{Flavor Discrimination}

At this point, for each hemisphere there is a list of selected tracks. 
For hemispheres with no selected secondary vertices the list is empty, 
otherwise it includes the tracks in the secondary vertices and any cascade 
tracks which have been attached. 
From this list several signatures can be computed to discriminate 
between bottom/charm/light event hemispheres. These are the 
invariant mass of the selected tracks corrected for missing $P_t$ ($M_{hem}$), 
the total momentum sum 
of the selected tracks ($P_{hem}$), the distance from the IP to the vertex
obtained by fitting all of the selected tracks ($D_{hem}$), and the total
number of selected tracks ($N_{hem}$). 

\begin{figure}
\caption{Distributions of flavor discrimination variables: 
(a) $M_{hem}$, (b) $P_{hem}$, (c) $D_{hem}$, (d) $N_{hem}$, 
(e) neural network output $y_{hem}$, (f) $P_{hem}$ vs. $M_{hem}$
correlation for $c$ jets (g) same for $b$ jets.}
\label{f:selnn}
\vspace{.2in}
\begin{center}
\epsfig{figure=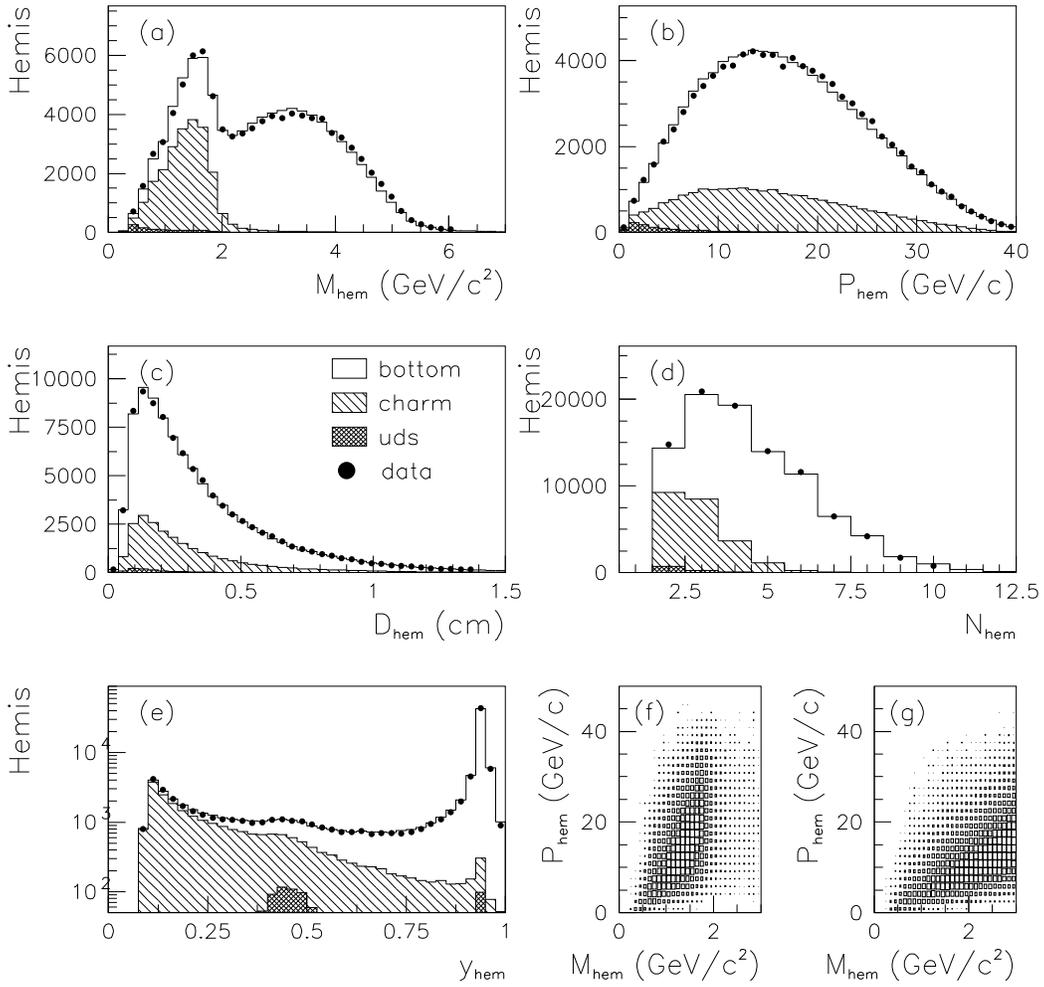,width=\textwidth}
\end{center}
\end{figure}

The four signatures given above are used as inputs for a neural network
trained to distinguish hemispheres in  bottom/charm/light events. The four
inputs and the neural network output $y_{hem}$ are shown in
Figure~\ref{f:selnn}. 
The $P_t$ corrected mass $M_{hem}$~\cite{sldrb}, is a particularly powerful 
discriminator to separate bottom and charm. For a detector with high
precision vertexing capability, decay vertices from $b$ and $c$ hadrons 
can both be very distinctively separated from the IP at high efficiency.
The large $b$ hadron mass is then the key kinematic information to allow 
separation of \bbbar\ and \ccbar\ with high purity, which is crucial
for these precision measurements. The raw vertex mass from the selected 
charged tracks can already give high purity $b$ tags once requiring mass 
above the charm threshold, but many $b$ decays with missing neutrals have 
an apparent low vertex mass. With the secondary vertex and PV positions 
very precisely measured at SLD, the $b$ hadron flight direction can be 
derived to compare with the vector sum of the selected secondary track 
momenta to estimate the missing $P_t$ with respect to the $b$ hadron flight 
direction. A corrected vertex mass is then calculated to compensate for 
the derived minimal missing mass, taking into account the vertex position 
errors: 
\begin{equation}
  M_{hem} = \sqrt{m_{c}^{2}+P_{t}^{2}} + |P_{t}|
 \label{eq:ptmass}
\end{equation}
where $m_c$ is the invariant mass of the tracks attached to the secondary
vertex in the hemisphere.

Vertices in $c$ quark jets, near the charm mass threshold typically have
small missing $P_t$, while many $b$ vertices near the charm mass threshold
receive a large missing $P_t$ correction to become well separated 
from \ccbar\ events. The missing particles for the low mass $b$ 
vertices generally lead to a lower apparent momentum calculated from 
the visible charged tracks so that the correlation between $P_{hem}$ 
and $M_{hem}$ presents another effective handle to further improve the
flavor separation, as shown in (f) and (g) of Figure~\ref{f:selnn},
comparing $c$ and $b$ jets.   
         
\begin{figure}
\caption{Purity vs. efficiency for hemispheres in 
(a) charm and (b) bottom events, 
as the selection neural network cut is varied.}
\label{f:effpur}
\begin{center}
\epsfig{figure=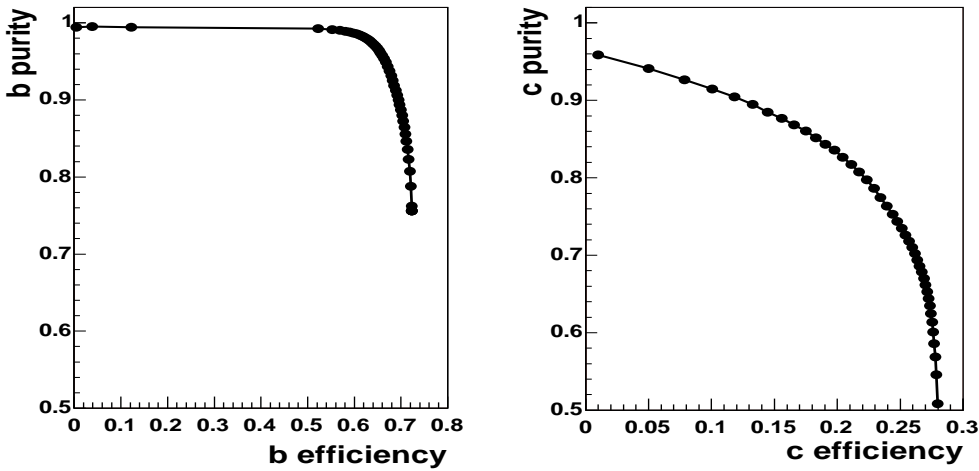,width=\textwidth,height=0.36\textheight}
\end{center}
\end{figure}

The flavor selection neural network is trained to put charm event 
hemispheres near $y_{hem}=0$, bottom event hemispheres near $y_{hem}=1$, 
and light-flavor background near $y_{hem}=0.5$. This allows a simple
selection of charm (bottom) event hemispheres by specifying an upper
(lower) limit for the output value $y_{hem}$. Figure~\ref{f:effpur} shows
the ranges of purity vs. efficiency which can be obtained for charm and
bottom event hemisphere tagging by adjusting this one cut.
As can be seen in Figures~\ref{f:vtxnn} and \ref{f:selnn}, the 
inputs used for the neural networks at various stages of the $y_{hem}$ 
tag construction are in reasonable agreement between data and MC. 
However, an exact agreement is not essential as the main tagging 
efficiencies will be measured directly from the data.        

\section{Measurement Method and Results}

Each event is divided into two hemispheres by the plane perpendicular
to the thrust axis. Both measurements apply flavor tags on each 
hemisphere separately to derive $R_b$ and $R_c$, and 
measure the major tagging efficiencies from data simultaneously.

\subsection{$R_b$ measurement} 

For the $R_b$ measurement, we define a single $b$-tag with
$y_{hem}>$0.75 and apply the classical double tag method as in
our previous measurement \cite{sldrb}. By counting the fraction
of tagged hemispheres, $F_s$, and the fraction of events with both
hemispheres tagged, $F_d$, we can measure $R_b$ by iteratively
solving the following equations:      
\bea
F_s & = &\effb R_b + \effc R_c + \effu (1-R_b-R_c),
\nonumber \\
F_d & = & C_b \efbb R_b + C_c \efcc R_c + \efuu (1-R_b-R_c).
\eea
where $\effu,\effc$ and $\effb$ are the hemisphere tagging
efficiencies for $uds$, $c$ and $b$ hemispheres, respectively,
and $C_q=\frac{\epsilon_q^{double}}{\epsilon_q^2}$,
are the hemisphere tag correlations for $b$ and $c$ events. 
We ignore the correlation for $uds$ events since we expect
0.23 double tagged events.
A Standard Model value of $R_c$=0.1723
is assumed. The dependence on MC simulation is greatly reduced 
by measuring \effb\ from data, while only the small \effc,\effu\
and correlations are estimated from MC. The measurement results and 
estimated MC parameters are tabulated in Table~\ref{tab:Rb-result}
for the 96 and 97-98 data samples. The errors are statistical only for 
both measured and MC parameters. The $R_b$ result has been
corrected by $-0.00149$ for hadronic event selection bias and by 
$+0.00024$ for $Z\gamma$ interference effect. 
  
\begin{table}[htb]
\begin{center}
\begin{tabular}{| l | rcl | rcl |}
\hline
            & \multicolumn{3}{|c|}{96}& \multicolumn{3}{|c|}{97-98}\\ 
\hline
$R_b$              & 0.21432&$\pm$&0.00289 & 0.21624&$\pm$&0.00104 \\
$\effb$(data)(\%)  &  56.63&$\pm$&0.67  &   62.01&$\pm$&0.24  \\ 
$\effb$(MC)(\%)    &  56.91&$\pm$&0.08  &   61.78&$\pm$&0.03  \\
$\Pi_b$ (\%)       &  97.8 &$\pm$&0.4     &  97.9&$\pm$&0.1   \\
$\effc$(MC)(\%)    &   1.24&$\pm$&0.02  &    1.19&$\pm$&0.01  \\ 
$\effu$(MC)(\%)    &  0.113&$\pm$&0.006 &   0.134&$\pm$&0.003 \\ 
$C_b-1$(MC)        & 0.0067&$\pm$&0.0015&-0.00012&$\pm$&0.00049\\
$C_c-1$(MC)        &   0.16&$\pm$&0.26  &    0.30&$\pm$&0.12 \\ 
N events           & 29996 &     &      &   191770&    &   \\
N single tag       &  3315 &     &      &   20738 &    &   \\
N double tag       &  2091 &     &      &   16048 &    &   \\
\hline 
\end{tabular}
\caption{$R_b$ result and tagging performance parameters. $\Pi_b$ is the
hemisphere $b$-tag purity. All errors are statistical only.}
\label{tab:Rb-result}
\end{center}
\end{table}   

\subsection{$R_c$ measurement}
 
Unlike the case of the $R_b$ measurement where the single $y_{hem}>0.75$ 
tag has already collected $b$ hemispheres with both high efficiency 
and high purity, the charm events are more spread out in the $y_{hem}$
distribution. Since the number of double tagged events
is very sensitive to the tag efficiency, the lower efficiency for charm 
tagging makes the use of different tags with quite different purities more 
profitable. We therefore use a multi-tag approach for the $R_c$ 
measurement.     
 
We divide hemispheres with a secondary vertex into four tag categories,
depending on the output value of the neural network, $y_{hem}$. 
A $b$ tag (tag~4) with $y_{hem} > 0.75$, a charm tag (tag~1) 
with $y_{hem} < 0.3$, a low-purity
$b$-tag (tag~3) with $0.5 < y_{hem} < 0.75$ and a low-purity charm tag (tag~2) 
with $ 0.3 < y_{hem} < 0.5 $. Hemispheres without a secondary vertex are classified
in the tag~0 category. In total we have 15 different event 
categories $N_{ij}, i\leq j \leq 4$ for the different tag combinations
with a predicted fraction of the number of events $G_{ij}$ :
\begin{equation}
G_{ij} = \kappa [ \epsilon_b^i \epsilon_b^j C_b^{ij} R_b
    + \epsilon_c^i  \epsilon_c^j C_c^{ij} R_c 
    + \epsilon_{uds}^i \epsilon_{uds}^j C_{uds}^{ij} (1-R_b-R_c) ]
  \label{Equ_Rcm} 
\end{equation}
with $\kappa=1$ for $i=j$ and 2 for $i \neq j$.
$\epsilon_q^i$ is the efficiency for quark $q$ to give a tag $i$.
$C^{ij}$ is the tag correlation between tag $i$ and tag $j$ similarly
defined as in the $R_b$ case.  
Since $N_{00}$ is determined by the total number of events
and the other event counts,
we have 14 independent equations for the event fractions 
$F_{ij} = N_{ij}/N_{\mbox{total}}$.

A small number of $uds$ events produce a secondary vertex.
This has two reasons. The first is gluon splitting to $b\overline b$
and $c \overline c$. The result is a real heavy quark of which the decay
is well modeled in our simulation. These events mainly populate the 
high purity $b$ and $c$ tag categories and 
the errors on the measured rates dominate the systematic uncertainty
assigned to this effect.
The other source is incorrect reconstruction in our detector. 
These events typically cluster around $y_{hem} = 0.5$.
This is less well modeled, and
in the next section we estimate the error on the rate predicted by our simulation 
to be 10\%. The $N_{02}$ and $N_{03}$ categories contain most of these 
events. To avoid a bias from these low purity bins,
we de-weight them by taking as the 
error on the event fraction, $\sigma_{F_{ij}}$, the binomial error on the 
bin contents and the systematic effect from varying the non-gluon 
splitting $uds$ efficiency by 10\%, added in quadrature.

We minimize 
\begin{equation}
  \chi^2 = \Sigma_{i,j, i\leq j} (F_{ij} - G_{ij})^2 / (\sigma_{F_{ij}})^2
\end{equation}
as a function of the following 9 parameters:
$R_b$, $R_c$, $\epsilon_b^j$ ($j=1,4$) and $\epsilon_c^j$ ($j=1,3$).
The $c$-quark efficiency for the $b$-tag, $\epsilon_c^4$, all light quark
efficiencies, $\epsilon_{uds}^j$ $(j=1,4)$ and the hemisphere correlations
$C^{ij}_q$ are taken from Monte Carlo. Only 
a few of the correlations are different from 1 in a statistically significant
way. The others are set to 1.

The $R_c$ fit results are summarized in Table~\ref{tab:effs}.  
The $R_c$ values have been corrected by $-0.0004$ for
$Z\gamma$ interference, the event selection bias is zero in simulation.
The $R_c$ values are given at a central $R_b$ 
value of $0.2157$. The measured value of $R_b$ agrees with the determination 
from the $R_b$ measurement. There is a good agreement between the efficiencies
in Monte Carlo and Data in the high purity tags 1 and 4. The efficiency for the
lower purity tags 2 and 3 for charm is higher in Monte Carlo than in the 
97-98 data sample. 
The Monte Carlo efficiency is quite sensitive to physics parameters like
the charmed hadron production fractions, their decay multiplicities and
their lifetimes. The measured value of the efficiency can be reproduced 
in the Monte Carlo, by varying some of these parameters 
within their allowed range as is done in the study of systematic uncertainties.
Since we extract the efficiencies from the data, the measured value of $R_c$ is
insensitive to these variations.

\begin{table}[htb]
\begin{center}
\begin{tabular}{| l | c c | c c |}
\hline
            & \multicolumn{2}{|c|}{96}& \multicolumn{2}{|c|}{97-98}\\ 
\hline
&  data & MC & data & MC\\
\hline
$R_c$ & $0.1678 \pm 0.0091$  & & $ 0.1752 \pm 0.0033$ &  \\
$\epsilon^1_b$ (\%) & $2.22 \pm 0.18$ & $2.01$ & 
                 $2.52 \pm 0.07$ & $2.36$ \\
$\epsilon^1_c$ (\%) & $16.09 \pm 0.90$ & $15.75$ & 
                 $17.95 \pm 0.36$ & $18.56$ \\
$\epsilon^1_{uds}$ (\%) &  & $0.088$ &  & $0.063$ \\
$\epsilon^2_b$ (\%) & $2.83 \pm 0.20$ & $2.56$ & 
                 $2.97 \pm 0.07$ & $3.00$ \\
$\epsilon^2_c$ (\%) & $4.93 \pm 0.36$ & $5.05$ & 
                 $5.02 \pm 0.13$ & $5.72$ \\
$\epsilon^2_{uds}$ (\%) &  & $0.117$ &  & $0.124$ \\
$\epsilon^3_b$ (\%) & $4.68 \pm 0.24$ & $4.40$ & 
                 $5.10 \pm 0.09$ & $4.98$ \\
$\epsilon^3_c$ (\%) & $2.46 \pm 0.27$ & $2.59$ & 
                 $2.23 \pm 0.10$ & $2.87$ \\
$\epsilon^3_{uds}$ (\%) &  & $0.150$ &  & $0.140$ \\
$\epsilon^4_b$ (\%) & $56.8 \pm 0.64$ & $56.9$ &        
                 $62.18 \pm 0.23$ & $61.78$ \\
$\epsilon^4_c$ (\%) &  & $1.23$ &  & $1.19$ \\
$\epsilon^4_{uds}$ (\%) &  & $0.113$ &  & $0.134$ \\
$C_c^{11} - 1$ & & -0.0068 & &  0.012 \\
$C_b^{14} - 1$ & &  0.0032 & & 0.0011 \\
$C_b^{34} - 1$ & &  0.0064 & & -0.0015 \\
$C_b^{44} - 1$ & &  0.0067& & -0.00012 \\
other $C_q^{ij} - 1$ & &  0& & 0 \\
$\chi^2/d.o.f.$ & $6.66/5$ &  & $7.05/5$ & \\
\hline 
\end{tabular}
\caption{$R_c$ results and tagging performance parameters. The
errors on the MC efficiencies are small compared to systematic
uncertainties and therefore omitted}
\label{tab:effs}
\end{center}
\end{table}   

\section{Systematic Errors and Cross-checks}
\label{sec:sys}

The systematic uncertainties on $R_b$ and $R_c$
result from a combination of detector related 
effects and physics uncertainties in the simulation which 
affect our estimates of $\effc$, $\effu$  and $C_q$ in the case of
$R_b$ and $\epsilon_4^c$, $\epsilon^{uds}_i$ and all correlations for $R_c$. 
These systematic uncertainties are listed in 
Table~\ref{tab:syst} for the 
combined results for 96 and 97-98 data,
which are analyzed separately initially. 

\subsection{Hemisphere correlation cross check} 
\label{sec:correlation-comp}

With the statistical precision down to the $<$0.5\% level for $R_b$ 
and the $<$2\% level for $R_c$, the systematic uncertainties of the 
small subtle effects of hemisphere tag correlation become
important. As the correlations have a variety of origins and
the evaluation of their uncertainties will be spread over many
subsections to follow, we will first discuss the correlation sources 
to establish an understanding of the magnitude of their effects. 
Primarily as a cross check to constrain possible missing systematic 
sources for the hemisphere correlations, we decompose the efficiency 
correlation of the $b$ and $c$ tags into a set of 
approximately independent 
components which represent the major known sources of correlation 
between the two hemispheres in the \bbbar\ and \ccbar\ MC events.

To focus on understanding the physics sources, we use the large 
uniform 97-98 MC sample without tracking efficiency and resolution 
corrections, to avoid the statistical fluctuations introduced 
by the tracking corrections. The overall tracking systematic effects
will be evaluated separately in section~\ref{sec:tracking-sys}.    
For the $R_c$ measurement, there are many correlations between
the different tags. We present a representative $c$-tag 
of $y_{hem}<0.4$, for similar set of effects as for the $b$-tag.    
For this study we need to identify the relevant kinematic and 
geometric variables to see how the tagging efficiencies depend
on them and how the two hemispheres correlate on these variables.

The primary vertex (PV) shared between the two hemispheres is
an obvious source of correlation. A misreconstructed PV 
results in a negative correlation if the 
displacement is along the thrust axis. 
Displacement of the PV transverse to the thrust axis
would tend to positively correlate the two hemispheres. The
small and stable SLD IP in the $xy$ view is an average 
beam position over many events which greatly reduces the 
chance of large PV displacement in $xy$, 
and restricts the remaining effect to be only through the 
IP $z$ coordinate, which is reconstructed event by event.  
The total PV effect is studied by simply comparing results using the 
reconstructed PV and using the MC truth PV. 

Another major source is geometric correlations. The two hemispheres
are exactly back to back by definition. 
Due the cylindrical geometry of the detector, larger $|\cos\theta|$ 
typically means more multiple scattering, worse tracking efficiency 
and resolution, worse radial alignment etc. This will
affect 
the hemisphere at $\cos\theta$ and the opposite hemisphere
at $-\cos\theta$ in the same way. We calculate
the tagging efficiency in bins of the hemisphere axis $\cos\theta$ bin
and construct a function $\epsilon(\cos\theta)$. The
$\cos\theta$ correlation component can then be estimated from
$$ C = \sum_i \epsilon(\cos\theta_i)\cdot\epsilon(-\cos\theta_i) 
   \,f(\cos\theta_i) / \overline{\epsilon}^2 $$  
where $i$ is the index of the $\cos\theta$ bin, $f$ is the 
fraction of all events in bin $i$ and $\overline{\epsilon}$ is the 
average tagging efficiency of all hemispheres.  

Although the cylindrical geometry of the detector ensures 
uniformity at first order in azimuth, local performance 
variation of detector elements can still introduce an effect.
Similar to the $\theta$ component analysis, the $\phi$ component
is estimated from 
$$ C = \sum_i \epsilon_{forward}(\phi_i)\cdot\epsilon_{backward}(\phi_i+\pi) 
   \,f(\phi_i) / \overline{\epsilon}^2 $$ 
where the efficiency parameterization is done for forward ($+z$) 
and backward ($-z$) hemispheres separately.   

Tagging correlation due to variation of detector performance in 
time also needs to be addressed. This is in fact the primary reason 
that the 97-98 data and 96 data are analyzed separately. 
There is a significant VXD3 performance variation during the 96 run,
and the\ Monte Carlo simulation reproduces
these time dependent effects in detail. 
The time dependent component is estimated from 
$$ C = \sum_i \epsilon^2(R_i) \,f(R_i) / \overline{\epsilon}^2 $$  
where $R_i$ is a group of runs adjacent in time, $f(R_i)$ is the
fraction of all events in this group of runs. 

The $b/c$-tag efficiencies have a significant dependence on 
the heavy hadron momentum ($P_H$) and its angle to the hemisphere 
axis ($\xi$). 
There are various causes, such as gluon radiation, which can 
result in correlations of the heavy hadron momenta and angular distributions. 
For the momentum correlation component, we parameterize the 
hemisphere tagging efficiency in a 2D grid of $\epsilon(P_H,\xi)$ then 
sum over all events $i$ for the correlation for hemispheres 
1 and 2:
$$C = \sum_i \epsilon(P^i_{H1},\xi^i_1)\cdot\epsilon(P^i_{H2},\xi^i_2)
      / \overline{\epsilon}^2 $$ 
This component is 
only evaluated for the normal cases where two heavy quarks are 
in opposite hemispheres. The joint effect of the heavy hadron 
momentum and thrust angle is $\sim$20\% larger than the effect 
from momentum alone.  
 
For the extreme case of hard gluon radiation, with two heavy quarks
recoiling into the same hemisphere opposite the hard gluon, 
we simply calculate the effect if including and 
excluding these events in the overall sample.  The magnitudes of the 
various components and their sum compared with the overall
correlations calculated from the MC hemisphere and double tag rates 
are shown in Table~\ref{tab:corr-comp}. 
\begin{table}[htb]
\begin{center} 
\begin{tabular}{|l|rr|rr|} 
\hline
  Component     & \multicolumn{2}{|c|}{$(C_{b-tag}-1)\times 10^5$}
                & \multicolumn{2}{|c|}{$(C_{c-tag}-1)\times 10^5$}  \\
                              & 97-98  &  96      & 97-98  &  96     \\
\hline 
Primary vertex                &   $+46$ &  $+13$  & $-480$ & $-640$  \\  
Geometrical correlation $\theta$& $+49$ &  $+60$  & $+120$ & $-120$  \\
Geometrical correlation $\phi$ &   $-4$ & $+212$  &  $-10$ & $+170$  \\
Time dependence               &    $11$ & $+434$  &  $+40$ & $+870$  \\  
$B/D$ momentum and thrust angle& $+107$ &  $+95$  & $+820$ & $+760$  \\
Hard gluon radiation          &   $-37$ &  $-23$  & $+510$ & $+410$  \\
\hline 
Component sum                 &  $+170$ & $+670$  &$+1000$ & $+1450$ \\
MC overall correlation        &   $+42$ & $+891$  & $+420$ & $+1510$ \\
MC statistical error          &$\pm 47$ &$\pm 113$&$\pm 390$&$\pm 1210$ \\ 
discrepancy                   &  $+128$ & $-121$  & $+580$ &   $-60$ \\        
\hline                         
\end{tabular} 
\caption{Hemisphere correlation component check results for $b$-tag 
and $c$-tag.}
\label{tab:corr-comp} 
\end{center}  
\end{table} 
An important observation is that the magnitude of all components are small, 
which is especially true for the $b$-tag. The change in the value of the 
correlation $C$ translates to the fractional error of $R_b$ and $R_c$. So
effects at typically 0.1\% (1\%) or less for the $b$($c$)-tag can be compared 
with the $R_b$ ($R_c$) statistical error of $\sim$0.5\% (2\%). This 
illustrates the importance of the large tagging efficiency in driving a 
much smaller correlation and reducing sensitivity to correlation uncertainties, 
which is even apparent when comparing the $b$-tag with the $c$-tag. 
In the limit of 
100\% tagging efficiency, the correlation becomes irrelevant.   

The 96 samples are not large enough to draw detailed conclusions
given the large MC statistical error, but the difference of the
overall correlation between 96 and 97-98 is well explained by the
known time and $\phi$ dependent effects in 96.  
For the more statistically significant test with the 97-98 sample,
there is a noticeable discrepancy between the $b$-tag component 
sum and the MC overall correlation. 
Although at least part of the discrepancy 
could be statistical for this approximate cross check, we take half of 
discrepancies from the 97-98 analysis for both the $b$-tag and $c$-tag
as an additional systematic error associated with possible missing sources.

\subsection{Physics systematics}  

The physics systematic errors are mostly assigned by reweighting the
nominal simulated distributions to an alternative set of distributions
which correspond to the world average measurements and uncertainties of 
the underlying MC physics parameters~\cite{heavy}. 

\subsubsection{\gluQQ\ effect} 
 
\begin{figure}
\begin{center}
\epsfig{figure=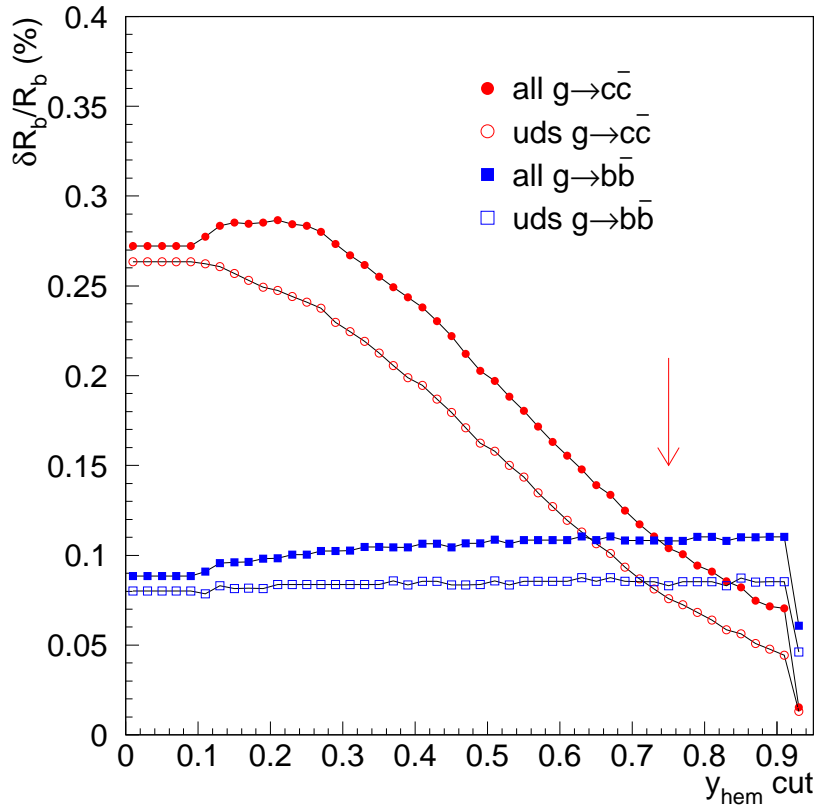,width=0.8\textwidth}
\caption{Components of the $R_b$ systematic error as a function of 
the flavor separation 
$y_{hem}$ cut for the various \gluQQ\ processes.} 
\label{fig:rbsys-gluqq}
\end{center}
\end{figure}

The \glubb\ and \glucc\ production rates are varied 
according to the experimental averages~\cite{heavy}. These measurement 
rates are significantly higher than the default JETSET MC as shown in
Table~\ref{tab:gQQ-gen}.
\begin{table} 
\begin{center}
\begin{tabular}{|l|ll|}  
\hline
                    &  \glucc\ (\%)   &  \glubb\ (\%) \\
\hline
  LEPEWWG standard  &   2.96$\pm$0.38 &  0.254$\pm$0.051 \\
  JETSET            &   1.357         &  0.142       \\
\hline  
\end{tabular} 
\caption{The world average measurements of \gluQQ\ compared to the
predictions of the JETSET generator.} 
\label{tab:gQQ-gen}
\end{center}
\end{table} 
The main sensitivity to the 
gluon splitting uncertainties is through the $uds$ tagging efficiencies.
Its effect on charm background under the $b$-tag is also noticeable.   
In the case of the $R_b$ measurement, we can see in detail 
in Fig.~\ref{fig:rbsys-gluqq} how  
\gluQQ\ affects the systematic uncertainty
on $R_b$ as a function of $y_{hem}$ cut. 

\subsubsection{$uds$ physics modeling} 

Among the fake $uds$ tags, a significant fraction involve 
$K^0_s$ and $\Lambda$ decay product. The MC generator level
$K^0_s$ and $\Lambda$ production are varied by $\pm 10\%$,
as recommended in \cite{heavy}, to derive the systematic 
uncertainty due to light hadron production.   

\subsubsection{charm production and decay modeling} 

The charm production and decay modeling affect the 
$R_b$ and $R_c$ measurements through the uncertainty on the small
charm tagging efficiency in the $b$-tag. They also affect the
$R_c$ measurement through the charm tag hemisphere correlations.
Many of the production and decay uncertainties only affect
charm tag correlations indirectly, e.g. through the primary 
vertex reconstruction.   

The different charmed hadrons have very different lifetimes and 
also rather different decay multiplicities so that the tagging 
efficiencies are also rather different. The $D^0,D^+,D_s$ and 
$\Lambda_c$ lifetimes are varied according to the errors on
their measured values as in PDG~\cite{pdg04}. 
The production fractions of these different
charmed hadrons are varied according to the recommendations 
of~\cite{heavy}.        

The tagging efficiency is in general higher for higher 
momentum charmed hadrons so that the energy spectrum of the 
charmed hadrons is another source of uncertainty. 
Charmed hadron fragmentation is studied by varying the 
average scaled energy in the range 
$\langle x_E \rangle=0.484\pm0.008$
using the Peterson fragmentation function~\cite{pete},
as well as by comparing the difference between the Peterson 
and Bowler models~\cite{bowler} for the same values of
$\langle x_E \rangle$. 

Charm tagging efficiency is sensitive to the
charged decay multiplicity. The fraction of decays with 
fewer than two tracks is a crucial source of 
inefficiencies. Higher multiplicity vertices are 
easier to identify, but the softer decay product momenta 
do not necessarily make the vertex resolution better.
The uncertainty due to $D$ decay charged multiplicities 
is estimated by varying each decay multiplicity fraction 
according to the Mark-III measurement errors ~\cite{mark3},
for $D^0,D^+$ and $D_s$ in turn, with a specific scheme 
as described in \cite{heavy}.
 
The production of $K^0$ in charm decays is another 
relevant source of uncertainty, although it may be somewhat
correlated to the charged multiplicity. In the case of 
$K^0_L$ or all neutral $K^0_s$ decays, it is a 
significant source 
of charged multiplicity loss. In the case of 
$K^0_s\ra \pi^+\pi^-$ decays, the decay product may affect 
the vertexing of charm decay differently depending on 
the $K^0_s$ decay length. There is no explicit recommended 
scheme from LEPEWWG. We reweight the 
$K^0$ multiplicity associated with charmed hadron decay
to correspond to the average multiplicity
range measured by Mark-III~\cite{mark3}.
 
Charmed hadron decays with fewer neutral particles have higher 
charged mass and are therefore more likely to be mistagged as a $b$.
Thus, an additional systematic uncertainty is estimated by
varying the rates of charmed hadron decays with no $\pi^0$s by 
$\pm 10\%$. This is an SLD specific estimate with particular
relevance to mass tags, which is typically not included
in LEP measurements.  

\subsubsection{$b$ production and decay modeling} 
\label{sec:bphys-sys}

The $b$ hadron production and decay modeling uncertainties only 
enter via the $b$-tag hemisphere correlation. Since the $B$ decays,
and to a large extent the $b$ jet fragmentations proceed independently
between the two hemispheres, most of these effects (apart from the 
QCD gluon radiation effect discussed separately in the next section)
enter indirectly through such effect as primary vertex reconstruction.  

The $b$-hadron lifetimes are varied by a typical current measurement
error of $\pm0.05$ps. We also vary the $\Lambda_b$ production ratio 
(not included in the LEPEWWG recommendation) by current measurement
uncertainty to acknowledge the fact that the $\Lambda_b$ lifetime is
significantly shorter than that of the $B$ mesons. 
$b$ hadron fragmentation is studied by varying the 
average scaled energy in the range 
$\langle x_E \rangle=0.702\pm0.008$
using the Peterson fragmentation function~\cite{pete},
as well as by comparing the difference between the Peterson 
and Bowler models~\cite{bowler} for the same values of
$\langle x_E \rangle$. The $B$ decay charged multiplicity 
distributions are reweighted to correspond to an average 
charged multiplicity uncertainty of $\pm$0.35. 
   
\subsubsection{Gluon radiation effects on tag correlations} 
\label{sec:gluon-sys}

Gluon radiation is a more direct source of correlation as it tends
to simultaneously lower the momenta of both heavy quarks as well
as changing their directions from the back to back topology. 
The tags are generally sensitive to heavy hadron momentum and 
their directions w.r.t. the thrust axis.  

We first discuss the extreme case of a very hard gluon radiation 
which causes two heavy hadrons to recoil
into the same hemisphere. This creates an anti-correlation of 
tagging efficiencies between the two hemispheres. 
Hard gluon radiation resulting in two $B$'s in the same hemisphere
occurs at a rate of 2.45\% in our simulation. This effect is reduced to
1.62\% in the \bbbar\ MC passing the analysis event selection.
In the case of the $b$-tag, the two $B$'s have
lower $B$ momentum and wider 
angles to the thrust axis, so that the $b$-tag efficiency is only 
slightly higher than a normal single $b$ hemisphere. 
The $c$-tag is more sensitive to the reduced $c$-hadron momentum
and the more confusing kinematic situation results in
a larger effect to the extent that the hemisphere with the two $c$-hadron 
even has a significantly lower efficiency than normal.   
For the actual systematic uncertainty, we follow 
the LEPEWWG recommendation to vary the MC rate by $\pm$30\%.
A cross check measurement (Appendix B.4.1 in \cite{wright-thesis}) of 
the rate of hard gluon radiation is performed with $b$-tags applied to 
each jet in 3 jet events. This analysis measured the ratio of 
same-hemisphere \bbbar\ rates between data and MC to be 0.82$\pm$0.09,
well within the $\pm$30\% variation.    
When the tagging efficiency of the hemispheres with two $B$'s is 
very close to the normal hemispheres with one $B$, there is a 
compensating effect between the double tag and hemisphere tag 
such that the overall correlation becomes very insensitive to 
the fraction of hard gluon radiation events. 
This results in a rather small error of $\delta R_b=-0.00002$.
In the case of $R_c$ analysis, the hard gluon events in both \bbbar\ and 
\ccbar\ are weighted up by 30\% for all tags and the combined effect 
of $\delta R_c=+0.00026$ is dominated by hard 
gluons in \ccbar\ events.

A systematic uncertainty is also assigned to the momentum
correlation between the two heavy hadron hemispheres, mostly due to gluon
radiation and fragmentation effects, which in turn translate to a
tagging efficiency correlation. In the case of the $B$ momentum 
correlation in \bbbar\ events, this is estimated by comparing
the $B$ momentum correlation in the HERWIG~\cite{herwig} and 
JETSET~\cite{jetset} event generators. At the parton level, all generators
give a similar correlation of $\sim$1.4\% between the $b$ quarks.   
At hadron level, the correlation coefficient for the $b$-hadron momenta 
in the two hemispheres is 1.55\% in JETSET, while the largest deviation
among different models is seen in HERWIG which gives up to +0.8\%
higher $B$ momentum correlation. 
We use half of this difference of 0.4\% as the variation to estimate 
the systematic error, according to the recommendation in \cite{heavy}. 
Given that different event selections can change 
the absolute correlation (excluding case of two $B$s in same hemisphere,
JETSET correlation reduces to 1.23\%, and applying $N_{jet}<4$ cut
further reduces this to 0.85\%), the ratio of 0.4\%/1.55\%=0.26 
is taken as the fractional uncertainty on the $b$ momentum correlation   
effect. 
A cross check analysis (Appendix B.4.2 in \cite{wright-thesis}) of $B$ 
hadron energy correlation is performed using the observed vertex 
momentum from charged tracks in double tagged events. The $B$
momentum correlation measured is verified to be in good agreement 
with MC to $\pm$20\%.
 
The angular distribution of $B$ flight direction and
thrust axis is also checked to be in good agreement between data and 
MC, where the $B$ direction is approximated by the line joining
$b$-tag vertex and PV.     
The component of the $b$ tagging correlation due to the $B$ momentum correlation
is estimated to correspond to $C_b-1=0.00107$, as is described
in section~\ref{sec:correlation-comp}, which translates to an error
on $R_b$ of 0.00006. 
There is no equivalent recommendation in \cite{heavy} for 
$c$-hadron momentum correlation in \ccbar\ events, mainly because 
this is only relevant for double tag analysis which is not done
at LEP. We similarly take $\pm$26\% of the $D$-momentum correlation
component in \ccbar\ correlation 
as a systematic uncertainty, which corresponds to 
$\delta C_{c-tag}=0.0020$ and $\delta R_c$ of 0.00022.

\subsection{Detector systematics} 
\label{sec:det-sys}

\subsubsection{Tracking resolution and efficiency}
\label{sec:tracking-sys}

The tracking resolution systematic effects, primarily due to residual 
detector misalignment are estimated from the observed shifts
in the track impact parameter distributions as a function of $\phi$
and $\theta$ in both $r$-$\phi$ and $r$-$z$ planes.
The typical impact parameter biases observed in the data are
$\sim2.5\mum$ ($\sim5\mum$) in the $r$-$\phi$ ($r$-$z$) plane.
A correction procedure is applied so that the MC tracks match the 
mean bias values of the data in various $(\phi,\theta)$ regions. 
This is a more realistic evaluation of alignment bias effects, 
where tracks passing the same detector region are biased in
a correlated manner. The actual corrections are implemented 
by dividing $\phi$ into 40 regions according to VXD3 ladder triplet 
boundaries and 4 sections in $\theta$. A detailed description of 
the resolution corrections can be found in 
\cite{chou-thesis}. Half of this correction is taken 
as the variation to evaluate the track resolution systematic 
uncertainty. 

\begin{figure}
\begin{center}
\epsfig{figure=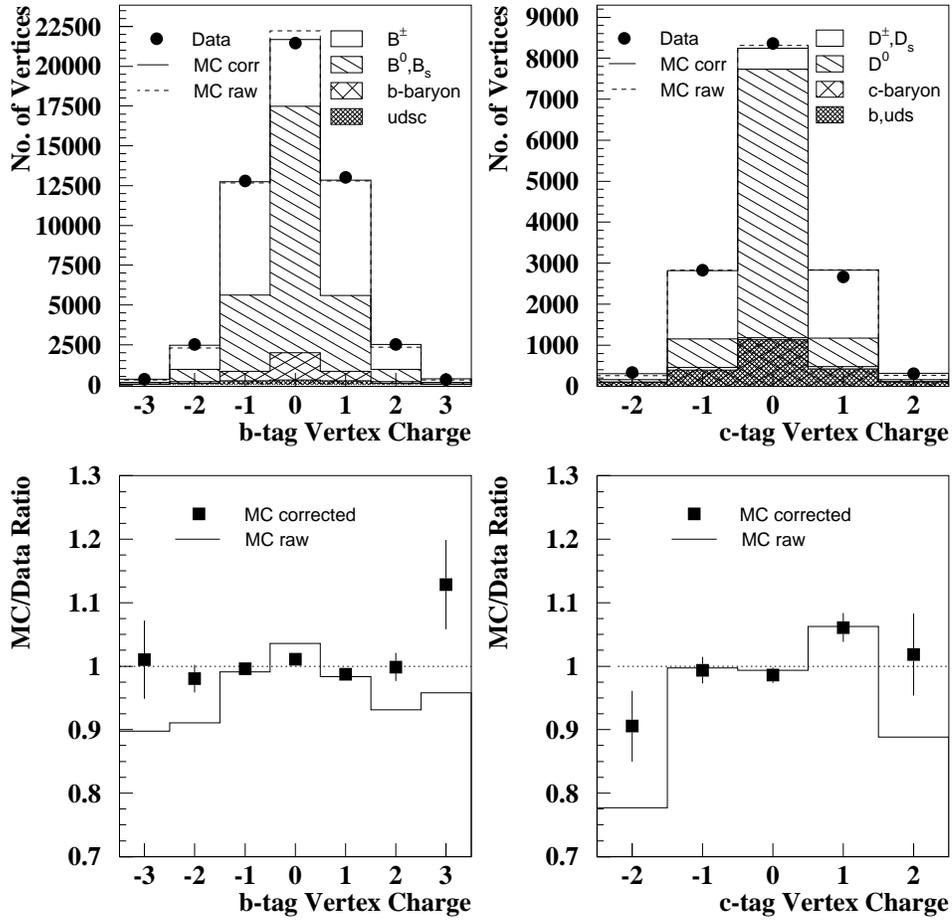,width=\textwidth}
\caption{Top: vertex charge distributions for $y_{hem}>0.75$ $b$-tag
(left) and $y_{hem}<0.4$ $c$-tag (right) hemispheres. The
data (dots) are compared with MC predictions with or without tracking corrections.
Bottom: the ratio of MC over data for before and after corrections} 
\label{fig:vtxq-detcorr}
\end{center}
\end{figure}

The uncertainty in the tracking efficiency is evaluated from a
comparison between data and MC for the fraction of all CDC tracks
which pass a set of quality cuts, and the fraction of good CDC tracks 
extrapolating close to the IP that does not have associated VXD hits. 
These studies indicate that the MC overestimates the tracking efficiency
by $\sim$1.5\% on average. A procedure for the random removal of tracks
in bins of $p_t$, $\phi$, and $\theta$ is used to correct the MC for 
this difference. 
Secondary vertex charge distributions are used as an independent 
check to verify that they agree
better between data and MC with these corrections applied, as shown
in Fig.~\ref{fig:vtxq-detcorr}.
This tests directly the relevant secondary vertex tracks 
independent of the modeling of fragmentation track multiplicity.
The $|Q|>1$ bins are particularly sensitive to tracking 
inefficiencies, while only weakly depend on heavy hadron decay 
multiplicity and production fractions.     

\subsubsection{IP tail}
Another potential systematic source is the possible tails in the 
IP position determination which can cause large tagging asymmetries. 
The \zmumu\ events have no indication of 
systematic tail effects. A further study 
\cite{walston-thesis} is done by examining the 
distribution of the distance between the hemisphere axis drawn 
through the fitted event primary vertex in $r-\phi$ to the IP for 
2-jet hadronic events with no secondary decay tracks. 

There is no significant discrepancy between data
and MC for 97-98 data, but there is a detectable tail in the 
96 data with $\sim$100\mum\ width for 0.5\% of the events. The full
effect of the tail is treated as a systematic uncertainty
for the 96 data.         

\subsubsection{Geometric and temporal correlation effects} 
\label{sec:geomcorr-sys}

Besides the issue of the uncertainty in the overall tracking efficiency,
MC modeling uncertainties on the efficiency nonuniformity due 
to local detector inefficiencies and time variations 
can lead to additional uncertainties on tagging correlations.  
The contribution of the $\theta$ and $\phi$ tagging efficiency 
dependence to the overall tagging efficiency correlation are 
estimated in section~\ref{sec:correlation-comp}. We estimate the 
uncertainty of the MC modeling of $\theta$ and $\phi$ 
dependency, by comparing the hemisphere tagging fractions 
between data and MC for the $y_{hem}>0.75$ $b$-tag 
and representative $y_{hem}<0.4$ $c$-tag. 
There is a generally good agreement between the data and MC within
the statistical errors. The only significant effect, the variation
in $\phi$ for 96 data, due to electronics failures for some of the
periods, is reasonably simulated by the MC. To 
estimate the correlation uncertainty, we simply take the ratio of
the all input hemisphere tagging fractions between data and MC for 
each $\phi$ and $\theta$ bin, to reweight the MC true signal tagging 
efficiency for the corresponding bin, as an approximated deviation.
The reweighted $\phi$ and $\theta$ dependent efficiencies are then
used to recalculate the tagging efficiency correlation components
as described in section~\ref{sec:correlation-comp}. The change
of the tagging correlation from the reweighting is taken as a 
systematic error. Note that correlation component calculations are 
insensitive to overall efficiency differences between data and MC 
which may enter in the reweighting, but only to shape
variations. The resulting systematic errors on the geometrical 
correlations are summarized in Table~\ref{tab:corrsys-geomtime}.
The correlation systematic error on $\Delta C_{b-tag}$ translates
directly to the fractional error $\delta R_b/R_b$,
and the error on the representative $c$-tag of $\Delta C_{c-tag}$ 
translates approximately to the fractional error $\delta R_c/R_c$.
The use of data vs MC ratio unfortunately introduces statistical
fluctuations on the estimated correlation change. A toy MC 
study indicates for example that the statistical fluctuation expected 
for the $\phi$ effect estimate is $\sim$0.00009 for 97-98
$b$-tag and as large as $\sim$0.0031 for the 96 $c$-tag. Some
of the changes are consistent with statistical fluctuation, but
we still conservatively take them as systematic uncertainties.    
      
\begin{table}
\begin{center} 
\begin{tabular}{|l|rr|rr|} 
\hline 
     & \multicolumn{2}{|c|}{$\Delta C_{b-tag}$}   
     & \multicolumn{2}{|c|}{$\Delta C_{c-tag}$} \\
     &   97-98  & 96   &  97-98 &  96    \\      
\hline 
$\theta$ effect   & 0.00016 & 0.00181 & 0.00043 & 0.00337 \\   
$\phi$ effect     & 0.00027 & 0.00037 & 0.00030 & 0.00409 \\
 time dependence  &       - & 0.00029 &      -  & 0.00055 \\
\hline
\end{tabular}
\caption{Correlation systematic uncertainties due to geometrical
and time dependent effects.}
\label{tab:corrsys-geomtime}
\end{center} 
\end{table}
    
For the time dependent effects on the tagging correlations there
is a very good uniformity in the 97-98 data. The smaller 96 
data sample does have significant time dependent effects, which
contribute to the overall uncertainties. 
The resulting systematic errors 
on the temporal correlations for the 96 analysis are summarized 
in Table~\ref{tab:corrsys-geomtime}.

\subsection{Event selection bias} 

The hadronic event selection flavor bias is evaluated from the 
MC for the basic hadronic event selection procedure and the last 
step of $N_{jet}<$4 cut separately. The event selection 
efficiencies and resulting bias on $R_b$ and $R_c$ are tabulated 
in Table~\ref{tab:evsel-bias}.
\begin{table} 
\begin{center} 
\begin{tabular}{|c|rr|r|}
\hline
 Stage          & Basic Had. Sel. &  $N_{jet}<4$  &  Total  \\
\hline    
$uds$ efficiency& 
$58.330\pm 0.050\%$ & $91.760\pm 0.037\%$ & $53.524\pm 0.051\%$ \\
$c$ efficiency& 
$58.592\pm 0.045\%$ & $91.695\pm 0.033\%$ & $53.725\pm 0.045\%$ \\
$b$ efficiency& 
$58.694\pm 0.036\%$ & $92.376\pm 0.025\%$ & $54.220\pm 0.036\%$ \\
\hline
$\Delta R_b$ & 
$0.00089\pm 0.00016$ & $ 0.00117\pm 0.00007$ & $0.00207\pm 0.00017$ \\
$\Delta R_c$ & 
$0.00040\pm 0.00018$ & $-0.00035\pm 0.00008$ & $0.00005\pm 0.00020$ \\
\hline 
\end{tabular} 
\caption{Event selection efficiencies and flavor bias for the 
uncorrected MC.}  
\label{tab:evsel-bias}  
\end{center} 
\end{table} 
The basic hadronic selection passes a slightly higher
fraction of \bbbar\ events, which 
is expected from the known higher charged multiplicity and other 
observed kinematic differences of $b$-jets compared to $uds$.
Given that the effect is only a few times the statistical error
$\sigma$, we simply take the 
MC statistical error as an uncertainty for this stage. The 
$N_{jet}<4$ cut on the other hand has a more statistically 
significant effect. The rate of \ccbar\ events passing
the $N_{jet}<4$ cut is consistent with that for $uds$
(most of the $\Delta R_c$ is actually the compensating effect
of $R_b$ bias). 

  The effect of the $b$ quark mass on the $\geq 4$ jet rate has 
significant theoretical uncertainties. The JETSET 7.4 MC used for 
our analysis is in fact known to have excessive suppression of
gluon radiation for heavy quarks \cite{norrbin-sjostrand}, when
comparing with data on 3 jet rates as used in the running $b$ 
quark mass measurements \cite{mbmz-meas}. We use the DELPHI 
measurement \cite{DELPHI-b4jet} of the ratio
$R_4^{bl}=f_4(\bbbar)/f_4(uds)$ to evaluate the event selection
bias in our analysis for the $<4$ jet requirement, where $f_4$ 
denotes the fraction of events with $\geq 4$ jets. The DELPHI measurement 
using the {\tt Cambridge} jet finder on all final state hadrons 
gives $R_4^{bl}=0.89\pm0.02$ at \ycut=0.006, with only very
small \ycut\ dependence. 
The overall 4 jet rate at this \ycut\ is very
similar to the 4 jet rates using JADE {\tt Yclus} jet finder 
on charged tracks with \ycut=0.02, as in our analysis.
We use the PYTHIA 6.228 generator \cite{PYTHIA62} as an 
intermediate reference to compute correction factors based on 
the ratios of ($1-R_4^{bl}$) with the same jet finding 
algorithms. We obtain a scaling factor of $S=0.50\pm0.13$ 
to be applied to the raw 4-jet cut bias in the JETSET 7.4 MC.
The uncertainty consists of equal contributions from the DELPHI 
measurement uncertainty and from the observed difference between 
using charged tracks and all final-state hadrons as input to the 
jet finder algorithms.  
Applying this scaling factor to the raw 4-jet bias of 
$\Delta R_b=0.00117$ in Table~\ref{tab:evsel-bias}, we obtain
a corrected 4-jet bias of $\Delta R_b=0.00059\pm0.00015$.  
The event selection bias for \ccbar\ is consistent with zero 
and therefore no correction is applied for $R_c$.

As a crosscheck to verify the $R_b$ event selection bias
estimate for the 4-jet cut, we also performed the analysis
without the $N_{jet}<4$ cut for the larger 97-98 data sample.
Taking into account the additional sensitivity to \gluQQ, 
we find the difference in measured $R_b$ 
due to the removal of $<4$-jet requirement is
$$ R_b ({\rm no\ <4\ jet\ cut}) - R_b ({\rm nominal\ analysis})
  = +0.00042 \pm 0.00030 \pm 0.00015 $$
where the first error is the uncorrelated statistical
uncertainty and the second error is the systematic error
on the 4-jet cut bias as estimated above. The difference is 
consistent with zero within statistics.

\subsection{Result stability checks} 

One possible source of systematic uncertainty
is poor modeling of the $uds$ background
that gives a secondary vertex due to badly reconstructed tracks.
As a cross-check we tried to extract this effect from the data.
We define a $uds$ tag by requiring no secondary vertex and no track with
a normalized 3D impact parameter of more than 2$\sigma$. This tag identifies
about 50\% of the $uds$, 15\% of the $c$ and 1.4\% of the $b$ hemispheres.
Adding this tag to the other ones, we fit for the same efficiencies as
before plus the ratio of $uds$ giving a secondary vertex in data over Monte
Carlo. We find a value of $1.1 \pm 0.1$ for this ratio. 

\begin{figure}
\begin{center}
\vspace{-2cm}
\epsfig{figure=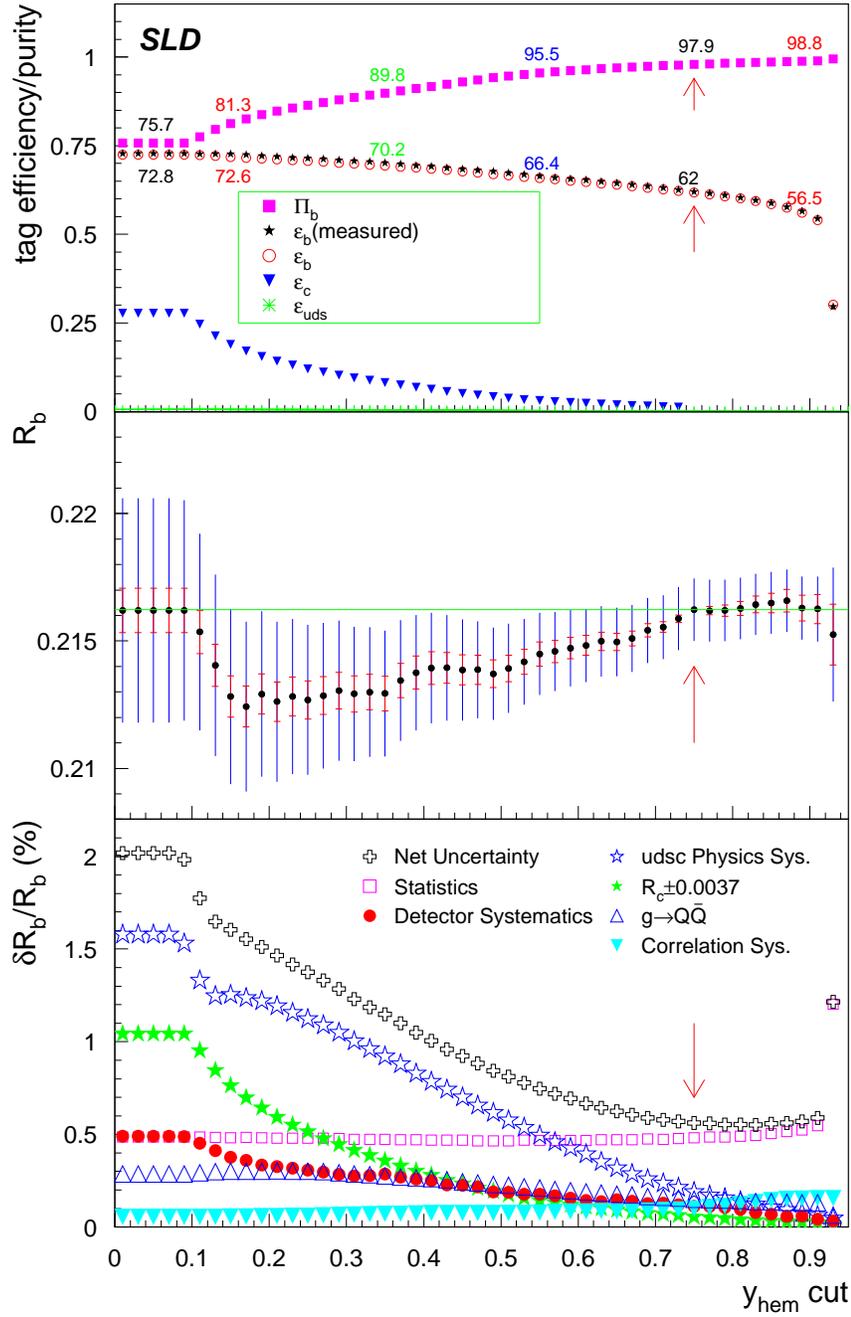,width=0.9\textwidth}
\caption{a) Tagging efficiencies and $b$ purity vs $y_{hem}$ cut. 
b) variation of the 
measured value of $R_b$, with the inner error bar being the uncorrelated
statistical error w.r.t. the nominal cut and the outer bar
total error. c) $R_b$
statistical and systematic error components as a function
of the $b$-tag selection cut, for the 97-8 analysis. The net uncertainty
does not include event selection uncertainties.}
\label{f:rbvar}
\end{center}
\end{figure}

We also examine the variation of the $R_b$ result as a function 
of the minimum $y_{hem}$ cut for the $b$ tag, over a wide range of 
$y_{hem}$ cut, as shown in Fig.~\ref{f:rbvar}. 
The $R_b$ 
result is stable within the $1\sigma$ total uncertainty envelope. 

Since the $R_c$ measurement uses the full range of $y_{hem}$ values,
there is no equivalent plot for this measurement. We do check the stability
of the result by moving the boundary between the high purity charm tag (1)
and the low purity charm tag (2) from 0.15 to 0.45 in increments of
0.05. We find that the $R_c$ result is stable within 0.0005.
The low purity tags showed some difference in efficiency for charm events
between data and Monte Carlo. Although we can reproduce the measured 
efficiency by varying some of the MC parameters in their allowed range
we perform an additional cross-check by removing the low-purity tag 3 
from the fit and re-measuring $R_c$. We find an insignificant
shift of $+0.0009 \pm 0.0014$ in 97-98 and $-0.0022 \pm 0.0032$ in 96.

\section{Conclusions}
\label{sec:concl}

We have measured the hadronic branching ratio of the $Z^0$ to $b$ 
quark and $c$ quark with our 96-98 dataset of $\sim$400,000 
hadronic \Z\ decays. The 96 and 97-98 results are combined,
with all common systematic uncertainties, including detector 
uncertainties, treated as fully correlated. The combined results are: \\

\begin{tabular}{lcllll}
  $R_b$ & = & \RBF & $\pm$\DRBSTAT {\mbox{(stat)}} 
            & $\pm$\DRBSYS {\mbox{(sys)}} & $\mp$\DRBRC ($R_c$) \\
  $R_c$ & = & \RCF & $\pm$\DRCSTAT {\mbox{(stat)}} 
            & $\pm$\DRCSYS {\mbox{(sys)}} & $\mp$\DRCRB ($R_b$) \\
\end{tabular}

\vspace{3mm} 
\par\noindent
For the $R_b$ measurement, combining this new measurement with 
our previously published result on the 93-95 data~\cite{sldrb}, 
we obtain:
\[
  R_b = 0.21594 \pm 0.00094 {\mbox{(stat)}} \pm 0.00074 {\mbox{(sys)}}
  \mp 0.00012 (R_c)
\]
The relative weights in the combined average for the 93-95:96:97-98 
(96:97-98) $R_b$ ($R_c$) measurements are 7:9:84 (11:89), dominated 
by the 97-98 result. These measurements are in good agreement with the 
Standard Model expectation of $R_b$=0.2156 (for $m_t$=178\gevcc) 
and $R_c$=0.1723. They can be compared with the 
average of LEP measurements \cite{CERN-EP-sum04} from a total of 
$\sim$16\,M hadronic \Z\ decays: \\
 
\begin{tabular}{lclll}
  $R_b$ & = & 0.21643 &$\pm$& 0.00073 \\
  $R_c$ & = & 0.1691  &$\pm$& 0.0047  \\
\end{tabular} 

\vspace{3mm}
\par\noindent
In conclusion, we have exploited the high resolution vertexing
capability and the small and the stable SLC IP for a precision test 
of Standard Model through the measurements of heavy quark production
fractions in \Z\ decays. Our new $R_c$ result is by itself more precise 
than the current world average \cite{pdg04}. These measurements confirm 
the Standard Model predictions at $\sim$0.6\% precision for $R_b$ 
and 2.1\% precision for $R_c$.

We thank the staff of the SLAC accelerator department for their
outstanding efforts on our behalf. This work was supported by the
U.S. Department of Energy and National Science Foundation, the UK Particle
Physics and Astronomy Research Council, the Istituto Nazionale di Fisica
Nucleare of Italy and the Japan-US Cooperative Research Project on High
Energy Physics.

\begin{table}[htb]
\begin{center}
\begin{tabular}{| l | r | r | }
\hline
 & {\bf $\delta R_b (10^{-5})$} & {\bf $\delta R_c (10^{-5})$}\\
\hline
MC statistics                     & 13  & 91\\
\glubb\ 0.254$\pm$0.051\%              & -24  & 9 \\
\glucc\ 2.96$\pm$0.38\%                & -23  & -101 \\
long lived light hadron production $\pm$10\%  & -1  & -1 \\
$D^+$ production  $0.233\pm0.028$      & -10 & -6 \\
$D_s$ production  $0.102\pm0.037$      & -11 & -15 \\
$c$-baryon production $0.065 \pm0.029$  &  -11 & 22 \\
charm fragmentation  & -18 & 18 \\
$D^0$ lifetime $0.415\pm0.004$ ps                    & -3 & 8 \\
$D^+$ lifetime $1.057\pm0.015$ ps                    & -2 & 5 \\
$D_s$ lifetime $0.467\pm0.017$ ps                    & -3 & -3 \\
$\Lambda_c$ lifetime $0.206\pm0.012$ ps              & -1 & -91 \\
$D$ decay multiplicity & -27 & 60 \\
$D$ decay $K^0$        & 19 & 56 \\
$D$ decay no-$\pi^0$        & -9 & 12 \\
$B$ lifetime $\pm0.05$~ps      & 0 & 5\\
$B$ decay   $\langle N_{ch} \rangle = 5.73\pm0.35$  & -20  & 3\\
$b$ fragmentation               & 4  & 26\\
$\Lambda_b$ production fraction $ 0.074 \pm0.030$    & 5  & -2\\
QCD hemisphere correlation          & 6  & 22\\
hard gluon radiation          & -2  & 26\\
tag geometry dependency    & 9  & 17\\
tag time dependency    & 1  & 1\\
component correlation             & 14  & 45\\
tracking resolution               & 27 & 22\\
tracking efficiency               & 13  & 3\\
$\langle IP \rangle_{xy}$ tail   & 2  & 0\\
event selection bias              & 17  & 20\\
4 jet rate in $b$ events 	& 15  & 0\\
\hline
{\bf $R_c=0.1723 \pm 0.0037$ }          & -12 &    \\
{\bf $R_b=0.2157 \pm 0.0010$ }          &  & -62  \\
\hline 
{\bf Total (excl. $R_{b/c}$)} & {\bf 73} & {\bf 200} \\
\hline 
\end{tabular}
\caption{ 
Summary of systematic uncertainties for $R_b$ and $R_c$.
A $-$ sign for an error means the value for $R_q$ goes
down when this parameter is varied upward. The errors are
assumed to be symmetric.}
\label{tab:syst}
\end{center}
\end{table}

\end{document}